\documentclass[letterpaper,twocolumn,10pt]{article}
\usepackage{usenix,epsfig,endnotes}
\usepackage{kotex}
\usepackage{multirow} 
\usepackage{colortbl}
\usepackage{color}
\usepackage{subfigure}
\usepackage{comment}
\usepackage{hyperref}

\begin{document}
\newcommand{\fsync}{\texttt{fsync()}}
\newcommand{\osync}{\texttt{osync()}}
\sloppy
\baselineskip=12bp
%don't want date printed
\date{}

%make title bold and 14 pt font (Latex default is non-bold, 16 pt)
\title{\Large \bf Barrier Enabled IO Stack for Flash Storage}

%for single author (just remove % characters)

\author{
{\rm Youjip Won}\\
Hanyang University
\and
{\rm Jaemin Jung}\\
Texas A\&M University
\and
{\rm Gyeongyeol Choi}\\
Hanyang University
\and
{\rm Joontaek Oh}\\
Hanyang University
\and
{\rm Seongbae Son}\\
Hanyang University
\and
{\rm Jooyoung Hwang}\\
Samsung Electronics
\and
{\rm Sangyeun Cho}\\
Samsung Electronics
}
\maketitle

% * <sangyeun@gmail.com> 2017-09-16T07:47:44.041Z:
% 
% We need to decide whether to use "IO" or "I/O" throughout this paper. I/O appears to be more standard, even in earlier papers of Prof. Won. I also suggest that we use the same term and style (e.g.,  Order Preserving ... or order preserving ...). It is more natural to use small caps unless it's absolutely needed to make a new unique noun.
% 
% ^ <sangyeun@gmail.com> 2017-09-16T07:49:28.245Z.
% Use the following at camera-ready time to suppress page numbers.
% Comment it out when you first submit the paper for review.
\thispagestyle{empty}
\subsection*{Abstract}

This work is dedicated to eliminating the overhead of guaranteeing the 
\emph{storage order} in modern IO stack. The existing block device adopts 
prohibitively expensive resort in ensuring  the  storage  order  among write requests: interleaving successive  write  requests  with transfer  and  flush. Exploiting the cache barrier command for the Flash
storage, we overhaul the IO scheduler, the dispatch module and the
filesystem so that these layers are orchestrated to preserve the ordering
condition imposed by the application till they  reach the storage surface.  Key ingredients of Barrier Enabled IO stack are \emph{Epoch based IO scheduling}, \emph{Order Preserving Dispatch}, and
\emph{Dual Mode Journaling}.  Barrier enabled IO stack successfully
eliminates the root cause of excessive overhead in enforcing the storage
order. Dual Mode Journaling in BarrierFS dedicates the
separate threads to effectively decouple the control plane and data plane of
the journal commit.
We implement Barrier Enabled IO Stack in server as well as in mobile
platform. SQLite performance increases by 270\% and  75\%, in server and in
smartphone, respectively. Relaxing the durability of a transaction,
SQLite performance and MySQL performance increases as much as by
73$\times$ and by 43$\times$, respectively, in server storage. 

\section{Motivation}
Modern IO stack is a collection of arbitration
layers;  IO scheduler, command queue manager, and storage writeback
cache manager.
Despite the compound uncertainties from the multiple layers of
arbitration, it is essential for the software writers to ensure
the order in which the data blocks are reflected to the 
storage surface, \emph{storage order},
e.g.~in guaranteeing the durability and the atomicity of a database 
transaction~\cite{mysql2007mysql, jeong2013stack, F2FS2015}, in  filesystem journaling~\cite{
tweedie1998journaling,  mathur2007new,sweeney1996scalability,JFS}, in 
soft-update \cite{mckusick1999soft,seltzer2000journaling}, or in copy-on-write or log-structure filesystems \cite{Rosenblum:1992:DIL:146941.146943,F2FS2015,rodeh2013btrfs,202319}.
\begin{figure}[t]
\begin{center}
\includegraphics[width=0.4\textwidth]{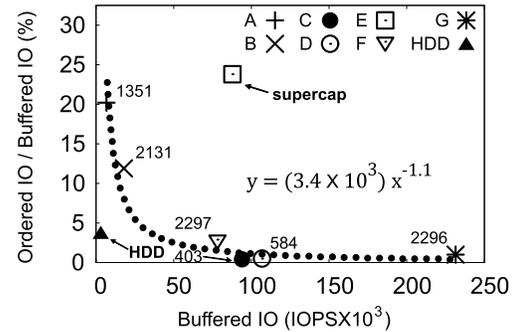}
\end{center}
\abovecaptionskip -2mm
\caption{Ordered write() IO vs.~Orderless write(), 
A: mobile/eMMC5.0, B:
mobile/UFS2.0, C: server/SATA3.0, D: server/NVMe, E:
server/SATA3.0 (supercap), F: server/PCIe, G: Flash array} 
\label{fig:B2F}
%\vspace{-5mm}
\end{figure}
Preserving the ordering requirement across the layers of the
arbitration is being achieved by an extremely expensive resort;
dispatching the following request only after the data block 
associated with the preceding request is completely transferred and
is made durable. We call this \emph{transfer-and-flush} mechanism.
For decades,
interleaving the writes with \emph{transfer-and-flush} has been
the fundamental principle to guarantee a storage order
in a set of requests \cite{req_flush,blockbarrier}. 

The concurrency and the parallelism in the Flash storage, e.g.~multi-channel/way
controller~\cite{Park2010parallelism,Chen2011parallelism}, large 
size storage cache~\cite{Narayanan:2008:WOP:1416944.1416949},
and deep command 
queue~\cite{dees2005native,ufs_standard,Xu:2015:PAN:2757667.2757684} 
have brought phenomenal performance improvement. 
State of the art NVMe SSD 
reportedly exhibits up to 750 KIOPS random read performance
~\cite{Xu:2015:PAN:2757667.2757684}, which is nearly 
4,000$\times$ 
of HDD's performance.  On the other hand, 
the time to program a Flash cell has barely improved if it has not  
deteriorated \cite{Grupp:2012:BFN:2208461.2208463}.
This is due to the adoption of the finer process (sub 10 nm)~\cite{helm201419,lee20167}, the multi-bits per cell (MLC, TLC, and QLC)~\cite{Chang:2015:ASP:2744769.2744790,cho2013adaptive} in 
the endless quest for higher storage density ~\cite{Mearian}.
Despite the splendid performance improvement of the Flash storage
claimed by the storage vendors, the service providers have difficulty
in fully utilizing the underlying high performance storage. 

Fig.~\ref{fig:B2F} alarms us an important trend.
We examine the  performance of write \emph{with} ordering guarantee
(\texttt{write()} followed by \texttt{fdatasync()}) against the one 
\emph{without} ordering guarantee (\texttt{write()}).
We test seven Flash storages with 
different degrees of parallelism.
In a single channel mobile storage 
for smartphone (SSD A),
the performance of ordered write is 20\% of that of the buffered write. In a thirty-two channel Flash array (SSD G), this ratio decreases to 1\%. 
In SSD with supercap (SSD E), 
the ordered write performance is 25\% of that of the
buffered write. There are two important observations. First, the overhead of 
transfer-and-flush becomes severe as the the degree of parallelism increases. Second,
use of Power-Loss Protection (PLP) hardware fail to eliminate the transfer-and-flush
overhead. 
The overhead is going to get worse
as the Flash storage employs higher degree of parallelism and denser Flash device.

Fair amount of works have been dedicated to address the overhead of storage order guarantee. The techniques deployed in 
the production platforms include non-volatile writeback
cache at the Flash storage \cite{guo2013low}, \texttt{no-barrier} mount 
option at the EXT4 filesystem \cite{barrierjournaling}, or transactional checksum 
\cite{IRON2005,kim2012tuning, asynccommit}.  Efforts as
transactional write at the filesystem 
\cite{Okun2002multiblock,Dabek2001CFS,Msync2013,F2FS2015,AdvFS2015} 
and transactional block device \cite{XFTL2013,Zhang2014versionFTL,min2015lightweight,ANVil2015,ouyang2011beyond} 
save the application from the overhead
of enforcing the storage order associated with filesystem journaling. A school of works address more fundamental aspects in 
controlling the storage order such as separating the ordering guarantee from durability
guarantee \cite{optfs2013}, providing a programming model to define
the ordering dependency among the set of writes \cite{gfs2007}, 
persisting a data block only when the result needs to be externally 
visible \cite{nightingale2006}.
These works share the same essential principle in controlling the storage order; transfer-and-flush. For example, OptFS\cite{optfs2013} checkpoints the data blocks only after the associated journal transaction becomes durable. Featherstitch\cite{gfs2007} realizes the ordering dependency between the \emph{patchgroups} via interleaving them with transfer-and-flush.

In this work, we revisit the issue of eliminating the transfer-and-flush overhead 
in modern IO stack. 
We aim at developing an IO stack
where the host can dispatch the following command before the data
blocks associated with the preceding command becomes durable and
 before the preceding command is
serviced and yet the host can enforce the storage order between them.

We develop a \emph{Barrier Enabled  IO stack} which
effectively addresses our design objective. 
Barrier enabled IO stack consists of the cache barrier-aware storage device, the 
order preserving block device layer and
the barrier enabled filesystem. Barrier enabled IO stack is built upon the
foundation that the host can control a certain partial order in which
the cache contents are flushed, \emph{persist order}.
Different from rotating media, the host can enforce a
persist order without the risk of getting anomalous delay in the
Flash storage. With reasonable
complexity, the storage controller can be made to flush the cache contents
satisfying a certain  ordering condition from the host \cite{ XFTL2013,prabhakaran2008transactional,lu2013lighttx}. 
The
mobile Flash storage standards already defines ``cache barrier'' command~\cite{emmc_standard} which precisely serves this purpose.
 For order preserving block device layer, the command dispatch mechanism and 
the IO scheduler of the block device layer 
are overhauled so that they can preserve partial order in
the incoming sequence of the requests in scheduling them. 
For barrier enabled filesystem, we define new interfaces, \texttt{fbarrier()} and 
\texttt{fdatabarrier()} 
to exploit the nature of order preserving block device layer. The 
\texttt{fbarrier()} and the \texttt{fdatabarrier()} system calls are the ordering guarantee only 
counter part of \texttt{fsync()} and \texttt{fdatasync()}, respectively. \texttt{fbarrier()} shares the same semantics as \texttt{osync()} of OptFS \cite{optfs2013}; it writes the dirty
pages, triggers filesystem journal commit and returns without persisting them. 
\texttt{fdatabarrier()} ensures the storage order between its preceding writes and
the following writes without flushing the writeback cache in between
and without waiting
for DMA completion of the preceding writes. It is a storage version of the  memory barrier, e.g.~\texttt{mfence} \cite{palanca2016mfence}. OptFS does not provide the one equivalent to \texttt{fdatabarrier()}. The order-preserving block device layer is filesystem-agnostic. We can implement \texttt{fbarrier()} and \texttt{fdatabarrier()} in any filesystems. We modify EXT4 to support \texttt{fbarrier()} and \texttt{fdatabarrier()}\footnote{The source codes are currently unavailable to public to abide by the double blind rule of the submission.
We plan to open-source it shortly.}. We only
present our result of EXT4 filesystem due to the space limit. We modify the journaling module of EXT4 and develop Dual Mode journaling
for order preserving block device. We call the modified version of  EXT4, the BarrierFS.

 Barrier Enabled IO stack not only removes the flush overhead but also
 the transfer overhead in enforcing the storage order. While large body  of the preceding
 works successfully eliminate the flush overhead, 
few works dealt with the overhead of  \emph{DMA transfer} in storage
 order guarantee.
The benefits of Barrier Enabled IO stack include
the following; 
\begin{itemize}
\itemsep0em
\item 
The application can control the storage order virtually without
any overheads; without being blocked or without stalling the queue.
\item  The latency of a journal 
commit  decreases significantly. The journaling module
can enforce the storage order between the journal logs and the journal commit mark
without interleaving them with flush and without interleaving them with DMA transfer.
\item Throughput of the filesystem journaling
improves significantly. Dual Mode journaling
 commits multiple transactions concurrently and yet can 
guarantee the durability of the individual journal commit.
\end{itemize}

Eliminating all the inefficiencies, the host now 
can successfully exploit
the concurrency and the parallelism in the underlying storage 
satisfying all ordering constraints. Relaxing the durability of a 
transaction,
SQLite performance and MySQL performance increase as much as by
73$\times$ and by 43$\times$, respectively, in server storage.

The rest of the paper is organized as follows. Section
\ref{section:background} introduces the background. 
Section
\ref{section:block}, section \ref{section:fs}, and section \ref{section:app}
explain the block device layer, the filesystem layer, and the
application of Barrier Enabled IO stack, respectively. Section
\ref{section:exp} and section \ref{section:rel} discusses the result of
the experiment and surveys the related works, respectively. Section
\ref{section:conc} concludes the paper.  

\vspace{-0.2cm}
\section{Background}
\vspace{-0.2cm}
\label{section:background}

\subsection{Orders in IO stack}

A write request travels a complicated route until the associated data 
blocks reach the storage surface.
The filesystem puts the  request to the IO scheduler queue. The 
block device driver removes one or more requests from the queue and 
constructs a command. It probes
the device and dispatches the command if the device is
\emph{available}. The device is \emph{available} if the command queue
at the storage device is not full. 
Arriving at the storage device, the command is
inserted into the command queue. The storage controller removes the
command from the command queue and services it, i.e.~transfers
the data block between the host and the storage. When the transfer
finishes, the device sends the completion signal to the
host. The contents of the writeback cache are committed to storage
surface either periodically or by an explicit request from the
host.

\begin{figure}[tp]
\centering
\includegraphics[width=0.95\linewidth]{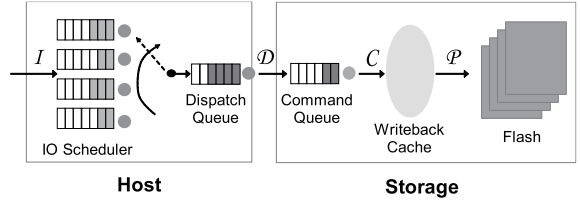}
\abovecaptionskip -2mm
\caption{Set of queues in the IO stack: the sources of
arbitration\label{fig:layers}} 
\end{figure}

 We define four types of orders in the IO stack; \emph{Issue Order},
$\cal I$,  \emph{Dispatch Order}, 
$\cal D$, \emph{Transfer Order},
$\cal C$, and \emph{Persist Order},
$\cal P$. The issue order
${\cal I}=\{i_1, i_2, \ldots, i_n\}$ 
is  a set of write requests 
issued by the application or by the
file system. The subscript denotes the
order in which the requests enter the IO scheduler.  The dispatch order
${\cal D}=\{d_1,d_2,\ldots,d_n\}$ denotes a set of the
write requests which are dispatched to the storage device. The subscript denotes the order in which the requests leaves the IO scheduler.  Transfer order,
${\cal C} = \{c_1, c_2, \dots, c_n\}$, is the set of transfer completions.
Persist Order ${\cal P} = \{p_1,p_2,\ldots,p_n\}$ is a set of operations which make the associated data blocks durable. Fig.~\ref{fig:layers} schematically 
illustrates the layers and the associated orders in the IO
stack. 
We say a certain partial order is preserved if the relative position
of the requests against a certain designated request, \emph{barrier},
are preserved.
We use the notation `=' to denote that
a certain partial order is preserved. 
\begin{comment}
Assume that there are the two dispatches
$d_k$ and $d_l$ in $\cal D$ and two 
transfers $c_{k'}$ and $c_{l'}$ in $\cal C$ each of which corresponds to $d_k$ and $d_l$, respectively.
We say $\cal D = C$ if  $\exists k$ and $\exists l$,  $(k < l) \mapsto  (k' \le  l')$ holds,
where $k$ and $l$ denote the index of a request in $\cal D$ and $k'$ and $l'$ denotes the index of the request in $\cal C$ corresponding to $d_k$ and $d_l$, respectively.
The IO scheduler can coalesce the two requests in the
issue order $\cal I$ into one request in the dispatch order $\cal D$. Two data blocks which have been transferred separately can
be made durable together atomically
if the two data blocks are programmed together to the same Flash page. Due to this mechanism, it is possible that $k' = l'$ while $k < l$. 
\end{comment}
We briefly summarize the source of arbitration
at each layer.
%\vspace{-.5 cm}
%
\begin{itemize}
\itemsep0em
\item ${\cal I} \neq {\cal D}$.  IO scheduler
reorders and coalesces the IO requests subject to their optimization criteria, e.g.~CFQ, DEADLINE, etc.  When there is no 
scheduling mechanism, e.g.~NO-OP scheduler \cite{axboe2004linux}
or NVMe \cite{NVMe}  interface, the dispatch order may  be equal to the issue order. 
\item
 ${\cal D} \neq {\cal C}$. Storage controller
 freely 
 schedules the commands in its command queue.
Also, the data blocks can be transferred out of order due 
to the errors, time-out and retry.
\item 
${\cal C} \neq {\cal P}$. The cache replacement algorithm, mapping table update algorithm, and storage controller's policy to schedule Flash operations governs the persist order independent
of the order in which the data blocks are transferred.
\end{itemize}
Due to the all these sources of arbitrations,
the modern IO stack is said to be \emph{orderless}~\cite{chidambaram2015orderless}.

\subsection{Transfer-and-Flush}
%
%\subsection{Transfer-and-Flush}
%
%{\color{red} yjwon: storage order 에 관한 내용이라서 이번 섹션이 좀더
%맞는 것 같읍니다.}
%

Enforcing a storage order corresponds to preserving a partial order between
issue order $\cal I$ and persist order $\cal P$, i.e.~satisfying the condition $\cal I = P$. It is equivalent to 
collectively
enforcing the individual ordering constraints between the layers; 
\begin{equation}
\cal (I = P) \equiv (I = D) \wedge (D = C) \wedge (C = P)
\label{eq:ip}
\end{equation}

%
%Despite that enforcing a storage order should be achieved through the collective
%efforts from multiple layers, the most critical
%participant, the storage device, does not provide  
%any mechanism to control the persist order. 
%
Modern IO stack has evolved under the assumption
that the host cannot control the persist order, i.e.~$\cal C \ne P$. 
\emph{Persist order} specifically denotes the order in which the
contents in the writeback cache are persisted whereas \emph{storage 
order} denotes an order in which the write 
requests from the filesystem are persisted. 
For rotating media such as hard disk drive, the disk 
scheduling is entirely left to the storage device due to its 
complicated  sector geometry hidden from outside \cite{gim2010extract}.
Blindly enforcing a certain persist order may bring unexpected delay in IO service.
Inability to control the persist order, $\cal C \ne P$, 
is a fundamental limitation of
the modern IO stack, which makes
the condition  $\cal I = P$ in Eq.~\ref{eq:ip} unsatisfiable.

To circumvent this limitation in satisfying a storage order, 
the host takes the indirect 
and expensive resort to satisfy each component in Eq.~\ref{eq:ip}.
First, after dispatching the write command to the storage device,
the caller is blocked until the associated DMA transfer
completes, \emph{Wait-on-Transfer}. This is to prohibit the storage 
controller from servicing the commands in out-of-order manner and
to satisfy the transfer order, $\cal D = C$. This may stall the 
command queue. When the DMA transfer completes, the caller issues
the flush command and blocks again waiting for its completion.
When the flush returns, the caller wakes up and issues the following
command; \emph{Wait-on-Flush}. These two are used in tandem leaving the caller under 
a number of context switches. 
Transfer-and-flush
is unfortunate sole resort in enforcing the storage order in a modern orderless IO stack.

\subsection{Analysis: \texttt{fsync()} in EXT4}
\begin{comment}
\begin{figure}[h]
\begin{center}
\includegraphics[width=0.45\textwidth]{./figure/flow_original_fsync.eps}
\end{center}
\caption{\texttt{fsync()} in EXT4}
\label{fig:ext4_ordered}
\end{figure}
\end{comment}
We examine how the EXT4 filesystem
controls the storage
order among the data blocks, journal descriptor, journal logs and journal 
commit block in \texttt{fsync()} in Ordered mode
journaling.
In Ordered  mode, EXT4 ensures that data blocks are persisted before the associated  journal transaction does.
\begin{comment}
\texttt{fsync()}  accounts for dominant fraction of IO's in popular workloads, e.g.~OLTP \cite{sysbench}, smartphone \cite{sqlitefamous, jeong2013stack} or mail server \cite{wilson2008new}. 
\end{comment}

\begin{figure}[btp]
\centering
\includegraphics[width=0.45\textwidth]{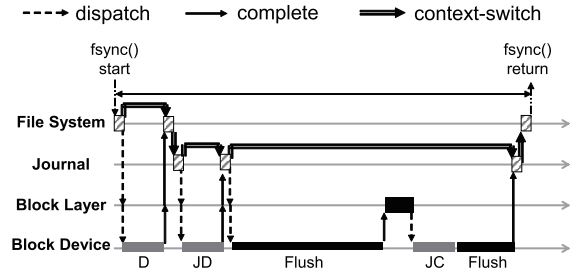}
\caption{DMA, flush and context switches in \texttt{fsync()}\label{fig:dma}}
\vspace{-.5cm}
\end{figure}

Fig.~\ref{fig:dma} illustrates the behavior of an 
\texttt{fsync()}. The application  dispatches the
write requests for the dirty pages, \emph{D}. 
After dispatching the write requests, the application 
blocks and waits for the completion of the associated
DMA transfer. When the DMA transfer completes, 
the application thread resumes and triggers the JBD thread to commit the
journal transaction. After triggering
the JBD thread, the application thread sleeps again.
When the JBD thread makes journal transaction 
durable, the \texttt{fsync()} returns, waking up the caller.
The JBD thread should be triggered
only after $D$ are completely. Otherwise, 
the storage controller may service the write requests for $D$, $JD$ and $JC$
in out-of-order manner  and storage controller
may persist the journal transaction prematurely before $D$ reaches the 
writeback cache. In this happens, the filesystem can be recovered 
incorrectly in case
of the unexpected system failure.

A journal transaction consists of the journal descriptor block,
one or more  log blocks and the journal commit block. 
A transaction is usually
written to the storage with two requests: one for
writing the coalesced chunk of the journal descriptor block and the log
blocks and the other  for writing the commit block. In the rest of
the paper, we 
will use \emph{JD} and \emph{JC} to denote the coalesced chunk of
the journal descriptor and the log blocks, and the commit block, respectively. 
JBD needs to enforce the storage order in two situations. $JD$ needs to be made durable
before $JC$. The journal transactions need to be made durable
in the order in which they have been committed. When any of the two conditions are violated, the file
system may recover incorrectly in case of unexpected system failure \cite{tweedie1998journaling,optfs2013}.
JBD interleaves the write request for $JD$ and the 
write request for $JC$ with transfer-and-flush. 
To control the storage order between the transactions,
JBD thread waits for $JC$ to become durable  before it starts
committing the next journal transaction. 

An \texttt{fsync()} can be represented as a tandem of Wait-on-transfer and Wait-on-flush as in Eq.~\ref{eq:fsync}.
$D$, $JD$ and $JC$ denote the write request 
for $D$, $JD$ and $JC$, respectively. `xfer' and `flush' denote wait-for-transfer  and wait-for-flush, respectively. 
\begin{equation}
\mbox{
$D$$\rightarrow$xfer$\rightarrow$$JD$$\rightarrow$xfer $\rightarrow$
$\underbrace{\mbox{flush$\rightarrow$$JC$$\rightarrow$xfer$\rightarrow$ flush}}_{\mbox{\texttt{FLUSH|FUA}}}$
}
\label{eq:fsync}
\end{equation}
In early days,
the block device layer was responsible for issuing the flush and for waiting
for its completion~\cite{Steigerward2007,iobarrier}.
%
\begin{comment}
\begin{figure}[tp]
\begin{center}
\includegraphics[width=0.45\textwidth]{./figure/flow_original_fsync.eps}
\end{center}
\caption{\texttt{fsync()} in EXT4}
\label{fig:ext4_ordered}
\end{figure}
\end{comment}
This approach blocks not only the caller 
but all the other requests which
share the same dispatch queue \cite{barrierjournaling}.
Since Linux 2.6.37 kernel, this role has been migrated 
from the block device
layer  to the 
filesystem layer~\cite{blockbarrier}. The filesystem uses flush option (\texttt{REQ\_FLUSH}) 
and force-unit-atomic option (\texttt{REQ\_FUA}) in writing
$JC$ and the filesystem blocks until it completes. With \texttt{FLUSH} 
option, the storage device flushes the writeback cache 
before servicing the command. With \texttt{FUA} option, 
the storage controller writes a 
given block directly to the storage surface. The last four steps in Eq.~\ref{eq:fsync} can be
compressed into a write request with \texttt{FLUSH|FUA} option.
\begin{comment}
In this approach, the \texttt{fsync()} can be modeled as in Eq.~\ref{eq:fsync2}.
%
\begin{equation}
\mbox{
$D$ $\rightarrow$ xfer $\rightarrow$ $JD$ $\rightarrow$ xfer $\rightarrow$
$\underbrace{JC_{\mbox{\texttt{FLUSH}$|$\texttt{FUA}}}}$
}
\label{eq:fsync2}
\end{equation}
\end{comment}
When the filesystem is responsible for waiting for the completion
of Flash, the other commands in the dispatch queue can 
progress after $JC_{\mbox{\texttt{FLUSH}$|$\texttt{FUA}}}$ is dispatched.
In both approaches, the caller is subject to transfer-and-flush overhead to   interleave  $JD$ and $JC$.

\section{Order Preserving Block Device Layer}
\label{section:block}
\vspace{-0.2cm}
\subsection{Design}
\label{section:bsio}
\begin{comment}
The order preserving block device layer is designed for
preserving the dispatch order, $\cal I = D$ and the transfer order, 
$\cal D = C$. It consists of order preserving IO scheduler
 and order preserving dispatch module. 
There are two design objectives: (i) to avoid 
imposing any unnecessary ordering constraints on the irrelevant requests
and (ii) to be seamlessly integrated with the existing scheduler.
\end{comment}

\begin{figure}[bp]
\begin{center}
\includegraphics[width=0.45\textwidth]{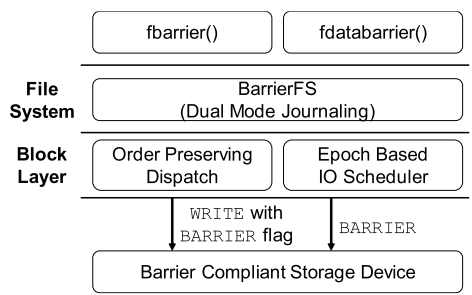}
\end{center}
\abovecaptionskip -2mm
\caption{Organization of the Barrier Enabled IO stack} 
\label{fig:overall}
\end{figure}

We overhaul the IO scheduler, the dispatch module and the write command
to satisfy each of three conditions, $\cal I = D$, $\cal D = C$, 
and $\cal C = P$, respectively.

In the legacy IO stack, the host has been entirely responsible for 
controlling the storage order; the host postpones sending the following command until 
it ensures that the result of the preceding command is made durable. 
In Barrier enabled IO stack, the host and the storage device share the responsibility. The host side block device layer is 
responsible for dispatching the commands 
in order. The host and the storage device 
collaborate with each other to transfer 
the data blocks (or to service the commands, equivalently) in order. 
The way in which the host and the storage device collaborate with each other will be detailed shortly. The storage device is responsible for making them durable in order. This effective orchestration between the
host and the storage device saves
the IO stack from the overhead of transfer-and-flush based
storage order guarantee.
Fig.~\ref{fig:overall} illustrates the
organization of Barrier Enabled IO stack.

The order preserving block device layer is
responsible for dispatching the commands
in order and for having them serviced 
in order. The IO scheduler and the command
dispatch module is redesigned to preserve
the order. 
Order preserving block device layer defines two types of write requests: 
\emph{orderless} and \emph{order-preserving}. 
There exists special type of order-preserving request called \emph{barrier}. 
We introduce two new attributes \texttt{REQ\_ORDERED} and  \texttt{REQ\_BARRIER} for the order-preserving request and the barrier request, respectively.
We call a set of order-preserving write requests which
can be reordered with each other as an \emph{epoch} \cite{condit2009better}. 
A
barrier request is used to delimit an epoch.

\subsection{\texttt{barrier write}, the command}

The ``cache barrier'', or ``barrier'' for short, command is defined 
in the standard command set for mobile Flash storage \cite{emmc_standard}. 
When the storage controller receives the barrier command, the controller guarantees 
that the data blocks transferred following the barrier command 
reach the storage surface after the data blocks transferred before the barrier 
command do without flushing the cache in between. A few eMMC products in the market 
support cache barrier command \cite{emmc_skhynix, emmc_toshiba}.
Via barrier command, the IO stack can satisfy the  persist order 
without cache flush. The essential condition $\cal C = P$ in ensuring 
the storage order can now be satisfied with the barrier command.

%A simplest way of exploiting the barrier command is to replace the existing flush command with the barrier command as in \cite{cachebarrier}. In this approach, however, the caller still needs to be blocked till the DMA transfer of $JD$ completes to dispatch the following command. Replacing the flush command with cache barrier command fails to fully exploit the potential benefit of "cache barrier", the ability to control the persist order. 

We start our effort with devising a more efficient barrier write command. Implementing a barrier as a separate command occupies one  
entry in the command queue and costs the host the latency of dispatching
a command. To avoid this overhead, we define  a barrier
as a command flag, \texttt{REQ\_BARRIER}, to the write command as in the case of \texttt{REQ\_FUA} 
or \texttt{REQ\_FLUSH}. 
In our implementation, we designate one  
unused bit  in the SCSI  command as a barrier flag. 

We discuss the implementation aspect of a barrier command. It is a matter of 
how the storage controller can enforce the persist order imposed by
the barrier command.
\begin{comment}
Efficient barrier write command support from Flash storage
device is the foundation of the barrier enabled I/O stack.
We begin our discussion of efficient barrier write command implementation
by describing a brief background of Flash storage device.
A Flash memory chip consists of multiple Flash blocks, each of which
contains multiple Flash pages. A Flash block should be erased before programming, programmed sequentially from its beginning page. A page size is typically 16KB, which may increase in upcoming new Flash memory generations.
Since Flash blocks require time consuming erase operations, FTL (Flash Translation Layer), a piece of software running on a storage 
controller, updates data in an out-of-place manner and maintains 
logical-to-physical mapping to keep track of valid data's location.
FTL is similar to the log-structured file system, maintains one or more
active ``log" blocks.
\end{comment}
When the Flash storage device has
Power Loss Protection (PLP) feature, e.g.~supercapacitor, supporting a barrier command is trivial.
Thanks to PLP, the writeback cache contents are always guaranteed to be durable. 
The storage controller can flush
the writeback cache in any order fully utilizing
its parallelism and yet can guarantee the persist order.
There is no performance overhead in enforcing the persist order.

For the devices without PLP, the barrier command can be supported  in three ways;
in-order write-back, transactional write-back or in-order recovery from crash.
In in-order write-back, the storage controller flushes  data blocks in epoch basis
and inserts some delay in between if
necessary. It may fail to fully exploit the
underlying parallelism in the storage controller.
 In transactional write,
the storage controller flushes the 
writeback cache contents as a single atomic unit \cite{prabhakaran2008transactional,lu2013lighttx}.
Since all epochs in the writeback cache are are flushed together, the constraint
imposed by the barrier
command is well satisfied. 
The performance overhead of transactional flush is 12\% in worst case with 
a traditional 
commit approach but can be eliminated by maintaining next page pointer at the spare area of the  Flash page~\cite{prabhakaran2008transactional}.

The in-order recovery method guarantees the persist order imposed by the
barrier command
through crash recovery routine. 
When multiple controller cores concurrently write 
the data blocks to multiple channels, one may have to use 
sophisticated crash recovery protocol such 
as ARIES protocol \cite{mohan1992aries} to recover the storage to consistent 
state. If the entire Flash storage is treated as a single log device, we can use simple crash recovery algorithm used in LFS \cite{Rosenblum:1992:DIL:146941.146943}.
Since the persist order is enforced by the crash recovery logic,
the controller is able to flush the writeback cache as if
there is no ordering dependency. The controller is saved
from  performance penalty at the cost of complexity in the recovery routine.

We implement the cache barrier command in UFS device, which is a commercial 
product used in the smartphone. We use simple LFS style recovery routine. 
The UFS controller  treats the entire
storage as a single log structured device and maintains an active segment in memory. FTL appends incoming data blocks to the active 
segment in the order in which they are transferred. It naturally satisfies the 
ordering constraints between the epochs. When an active segment becomes full, it 
is striped across the multiple Flash chips in log-structured manner.
In crash recovery, the UFS controller locates the beginning of the most recently
flushed segment. It scans the pages in the segment from the beginning
till it first encounters the page which has not been programmed properly.
The storage controller discards the rest of the  pages including 
the incomplete one.

Developing a sophisticated barrier-aware SSD controller is
subject to a number of design choices and
should be dealt with in detail in separate context. 
Through this work, we
demonstrate that  the  performance benefit in using the
cache barrier command  deserve the complexity of implementing it 
if the host side IO stack can properly exploit it.

%\vspace{-.5cm}
\subsection{Epoch Based IO scheduling}
%\begin{figure}[tbp]
%\begin{center}
%\includegraphics[width=0.45\textwidth]{./figure/epoch_manage.eps}
%\end{center}
%\caption{Epoch Management}
%\label{fig:jbd2cp}
%\end{figure}
%
%REQ_ORDERED
%RED_BARRIER
%

%

There are three scheduling principles
in Epoch based IO scheduling.
First, it preserves the partial order 
between the epochs. Second, the requests within an epoch can be 
freely scheduled with each other.
Third, the orderless requests can be scheduled freely
across the epochs. It satisfies  $\cal I = D$ condition.

The Epoch Based IO scheduler uses
existing IO scheduler, e.g.~CFQ, NO-OP and etc., to schedule the IO requests 
within an epoch. 
The key ingredient of the Order Preserving IO scheduler
is \emph{Epoch based barrier reassignment}. When the IO request enters
the scheduler queue, the order preserving IO scheduler examines if it is
a barrier request. If the request is not a barrier request, it is inserted as normal requests.
If the request is a barrier write request, IO scheduler removes
the barrier flag from the request and inserts it to the queue.
After the scheduler inserts a barrier write, the scheduler 
stops accepting more requests. The IO scheduler re-orders and 
merges the IO requests in the queue based upon its own scheduling discipline
e.g. FIFO, SCAN, CFQ. The requests in the queue either are
orderless or belong to the same epoch. Therefore, they can be freely scheduled with
each other without violating the ordering condition.
The merged request will be order-preserving if one of the 
constituents is order-preserving.
The IO scheduler designates the order-preserving request 
that leaves the queue last as a new barrier. This mechanism 
is called \emph{Epoch Based Barrier Reassignment}.
When there is no more order-preserving requests in the 
queue, the IO scheduler starts accepting the IO requests.
When the IO scheduler unblocks the queue, there can be one or more orderless requests
in the queue. These orderless requests can be scheduled
with the other requests in the following epoch. 
Differentiating
the order-preserving requests from orderless ones,
we avoid imposing unnecessary ordering constraint on the requests.
Currently, the Epoch based IO scheduler is implemented on top of existing
CFQ scheduler. Each process defines its own scheduler queue.

\begin{figure}[t]
\centering
\begin{comment}
\subfigure[Violation of Ordering Constraints\label{fig:scheduler}]
{\includegraphics[width=0.4\textwidth]{./figure/epoch_manage1.eps}} 
\\
\end{comment}
%\subfigure[Epoch Based Barrier %Reassignment\label{fig:scheduler_epoch}]
%{
\includegraphics[width=0.45\textwidth]{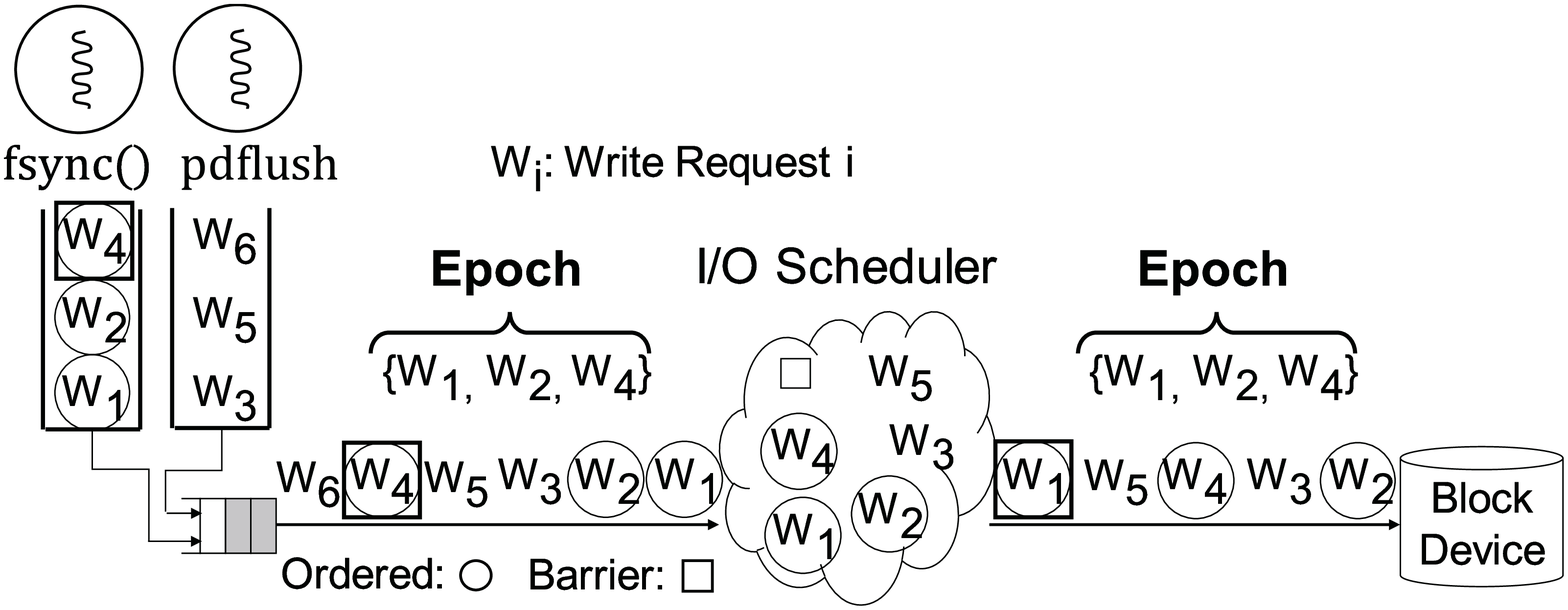}
%}
\caption{Epoch Based Barrier Reassignment
\label{fig:epoch}} 
\vspace{-.5cm}
\end{figure}

Fig.~\ref{fig:epoch} illustrates how the  barrier
reassignment works. The circular and the rectangular write request denote the order-preserving attribute and barrier
attribute, respectively. In Fig.~\ref{fig:epoch}, the application calls \texttt{fsync()} and in the  
mean time, \emph{pdflush} daemon flushes the dirty pages. In Fig.~\ref{fig:epoch},  \texttt{fsync()}
creates three write requests: $w_1, w_2$ and $w_4$. 
The filesystem marks the three requests as ordering preserving ones. 
The filesystem designates the last request, $w_4$, as a barrier write. \texttt{pdflush} creates three write requests $w_3, w_5$ and $w_6$. They
are all orderless. The requests from the two threads are fed to 
the IO scheduler with as $w_1, w_2, w_3, w_5, w_4^{barrier}, 
w_6$ in order. When the barrier write, 
$w_4$, enters the queue, the scheduler stops accepting the new request. 
There are only five requests in the queue, $w_1, w_2, w_3, w_4$ and $w_5$. $w_6$ 
cannot be inserted at the queue since the queue is blocked. 
The IO scheduler reorders the them and dispatches them in $w_2 w_3 w_4 w_5 w_1$ order. After they are scheduled, $w_1$ leaves the queue last.
The IO scheduler puts the barrier flag to $w_1$. 
In this scenario, the request $w_6$ is going to be scheduled with the requests in the 
following epoch.

\subsection{Order Preserving Dispatch}
\begin{figure}[t]
\centering
\subfigure[When Device is Available\label{fig:wod_avail}]{
   \includegraphics[width=0.4\textwidth]{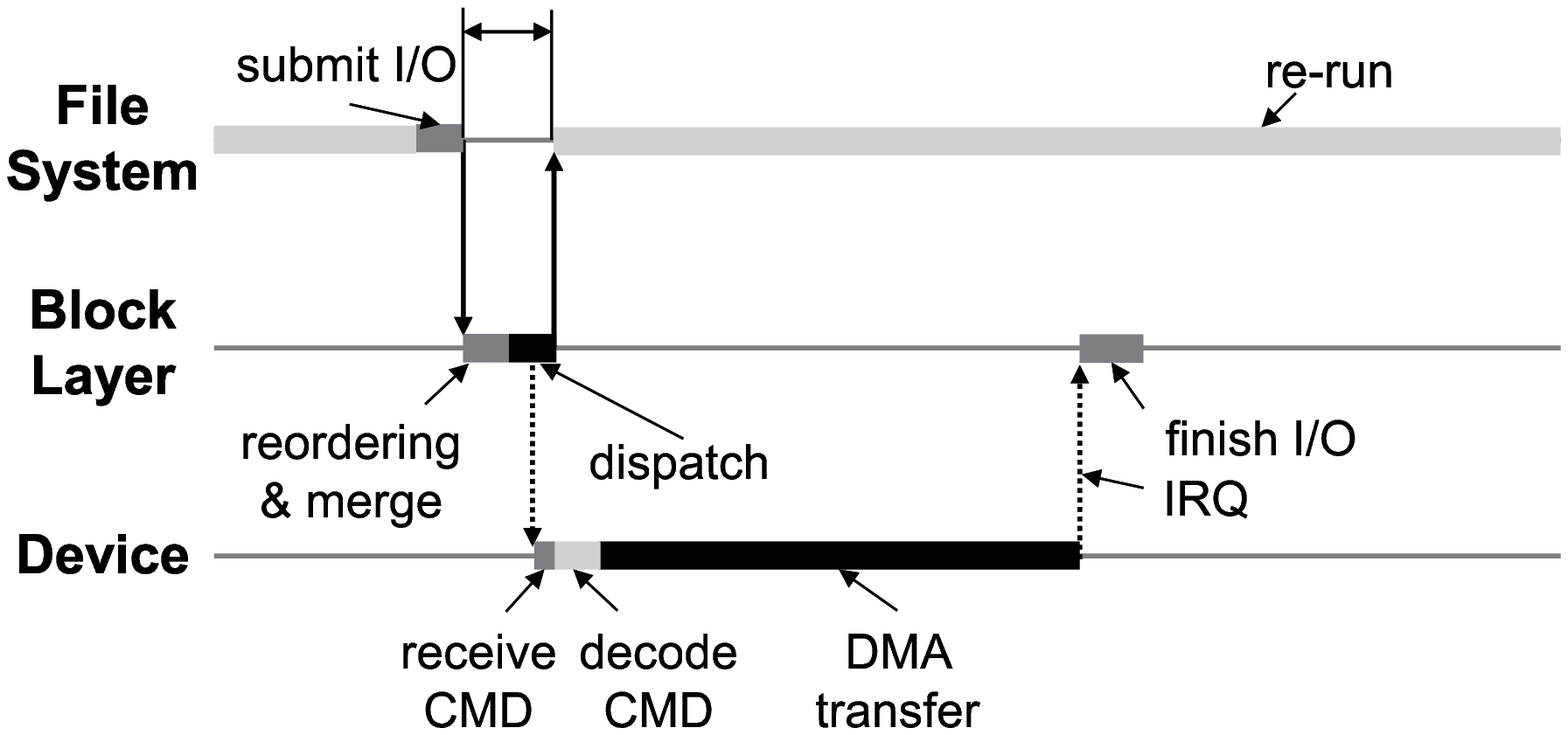}
   }
   \\ 
\subfigure[When Device is Busy\label{fig:wod_busy}]{
   \includegraphics[width=0.4\textwidth] 
   {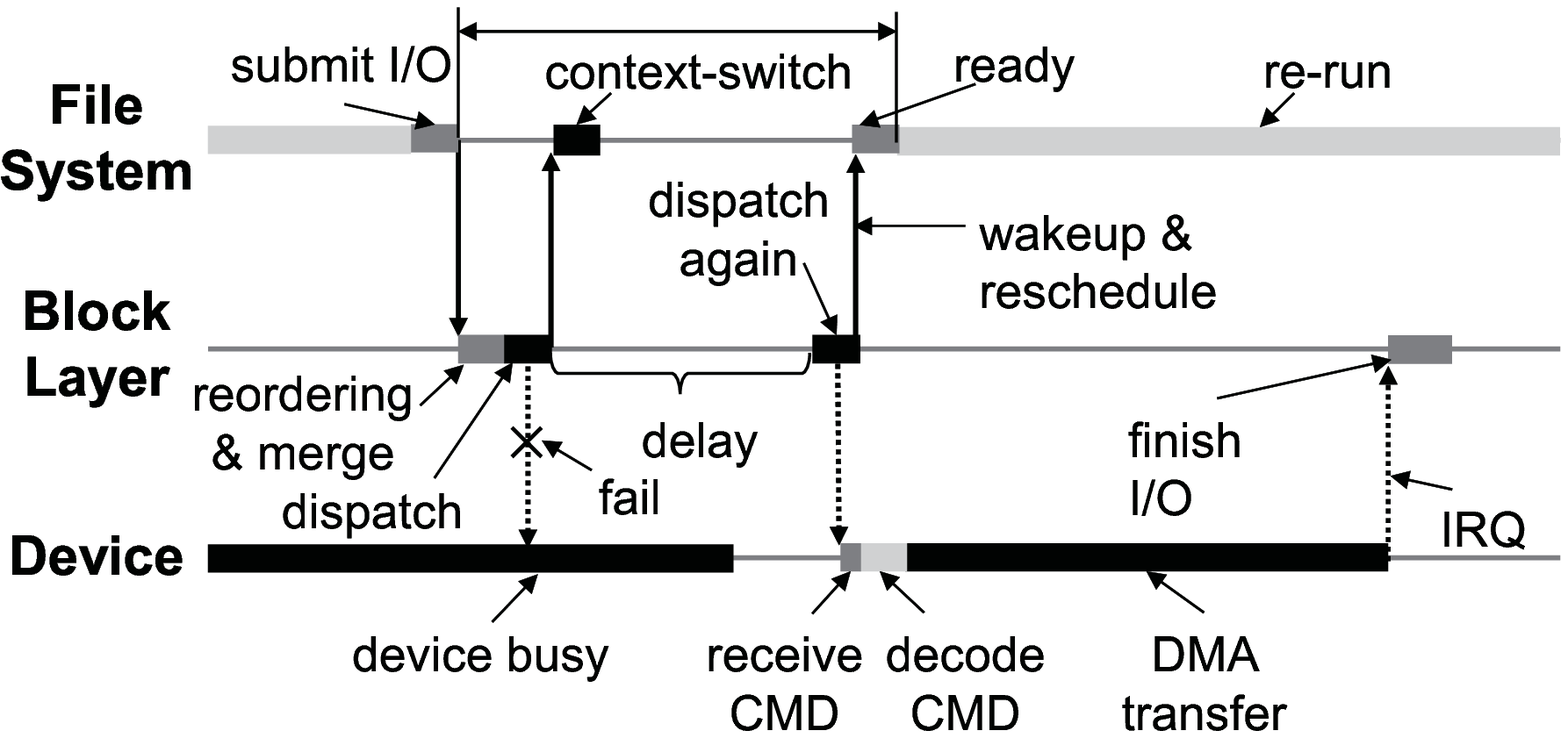}}
\caption{Order Preserving Dispatch\label{fig:dispatch}}
\vspace{-.5cm}
\end{figure}

The order preserving dispatch 
is a fundamental innovation of this work. 
In order preserving
dispatch, the host dispatches the following write request
when the storage
device acknowledges that the preceding request has successfully
been received (\ref{fig:wod_avail}) and yet the transfer order between
the two requests are preserved, i.e.~$\cal D = C$. The order preserving dispatch guarantees the transfer order without blocking the caller. Legacy IO stack 
controls the transfer order with 
\emph{Wait-On-Transfer}.
Wait-On-Transfer not only exposes
the caller to the context switch overhead
but also makes the IO latency less 
predictable. It may stall the storage device 
since the caller postpones dispatching the 
following command till the preceding command 
is serviced. Order preserving dispatch
eliminates all these overheads.

For order preserving dispatch, the only 
thing the host block device driver
does is to set the priority of a barrier 
write command to \emph{ordered} when 
dispatching it. 
Then, the SCSI compliant storage device 
automatically guarantees the transfer
order constraint in serving the requests. 
SCSI standard defines three command priority levels: \emph{head of the queue}, \emph{ordered}, and \emph{simple}\cite{scsicmd}, with which the incoming command is put at
the head of the command queue, tail of the command queue or at arbitrary position determined by the storage controller. In addition, the simple command cannot be inserted in front of the existing "ordered" or "head of the queue" commands. The \emph{head of the queue} priority
is used when a command requires an immediate service, e.g.~flush command. 
Via setting the priority of barrier write command
to \emph{ordered}, the host ensures the the data blocks associated with the write requests in the 
preceding epoch are transferred ahead of the data blocks associated with the barrier write. Likewise, the data blocks associated with the following epoch are transferred
after the data blocks associated with the barrier write is transferred. The transfer order condition is satisfied.

The caller may be blocked after dispatching the write request. This can happen when the
device is unavailable or the caller is
switched out involuntarily, e.g.~time quantum expires. For both cases,
the block device driver of the order preserving dispatch module uses the same error handling routine
adopted by the existing block device driver;
the kernel daemon inherits
the task and retries dispatching the request
after a certain time interval, 
e.g.,~3 msec for SCSI device~\cite{scsicmd} (Fig.~\ref{fig:wod_busy}).  The thread resumes once the request is dispatched
successfully.

\section{\texttt{BarrierFS}: Barrier Enabled Filesystem}
\label{section:fs}
\vspace{-0.2cm}
\subsection{ Programming Model}
We propose two new filesystem interfaces, \texttt{fbarrier()} and \texttt{fdatabarrier()} which
are the ordering guarantee only counter part to
\texttt{fsync()} and \texttt{fdatasync()}, respectively.
\texttt{fbarrier()} shares the same semantics with
\texttt{osync()} in OptFS \cite{optfs2013}. 
The salient feature of BarrierFS is \texttt{fdatabarrier()}. 
\texttt{fdatabarrier()} returns
after dispatching the write requests for dirty pages. 
With 
\texttt{fdatabarrier()}, the application can enforce
a storage order virtually without any overhead; without flush,
without waiting for DMA completion and even without context switch.
The following codelet illustrates the usage of the \texttt{fdatabarrier()}.
\begin{verbatim}
write(fileA, "Hello") ; 
fdatabarrier(fileA) ; 
write(fileA, "World")} 
\end{verbatim}
It ensures that ``Hello'' is written to the storage surface ahead of ``World''. 
Modern applications have been using expensive
\texttt{fdatasync()} to guarantee both durability
and ordering.
For example, SQLite which is the default DBMS in mobile device, such as Android,
iOS or Tizen uses \texttt{fdatasync()} to ensure that
the updated database node reach the disk surface ahead 
of the updated database header. In SQLite,
\texttt{fdatabarrier()} can  replace the 
\texttt{fdatasync()} when it is used for ensuring the 
storage order, not the durability.

The Barrier Enabled IO stack is filesystem agnostic.
\texttt{fbarrier()} and \texttt{fdatabarrier()} can be implemented
in any filesystem using proposed order preserving block device layer.
As a seminal work, we modify the EXT4 filesystem for order preserving block 
device layer. We optimize \texttt{fsync()} and \texttt{fdatasync()} for
order preserving block device layer and newly implement \texttt{fbarrier()} and \texttt{fdatabarrier()}.
We name the modified EXT4 as BarrierFS. \texttt{fbarrier()} in BarrierFS supports all journal modes in EXT4; WRITEBACK, ORDERED and DATA.
\subsection{Dual Mode Journaling}
\begin{comment}
\begin{figure}[ht]
\centering
{\includegraphics[width=0.45\textwidth]{./figure/BFS_cs.eps}}  
\caption{Concurrent Journaling, D: DMA for
dirty pages, JD: DMA for journal descriptor, JC: DMA for journal commit block
\label{fig:BFS_cs}}
\end{figure}
\end{comment}

\begin{figure}[ht]
\centering
\subfigure[\texttt{fsync()} in EXT4 with \texttt{FLUSH/FUA}\label{fig:EXT4_cs}]{\includegraphics[width=0.45\textwidth]{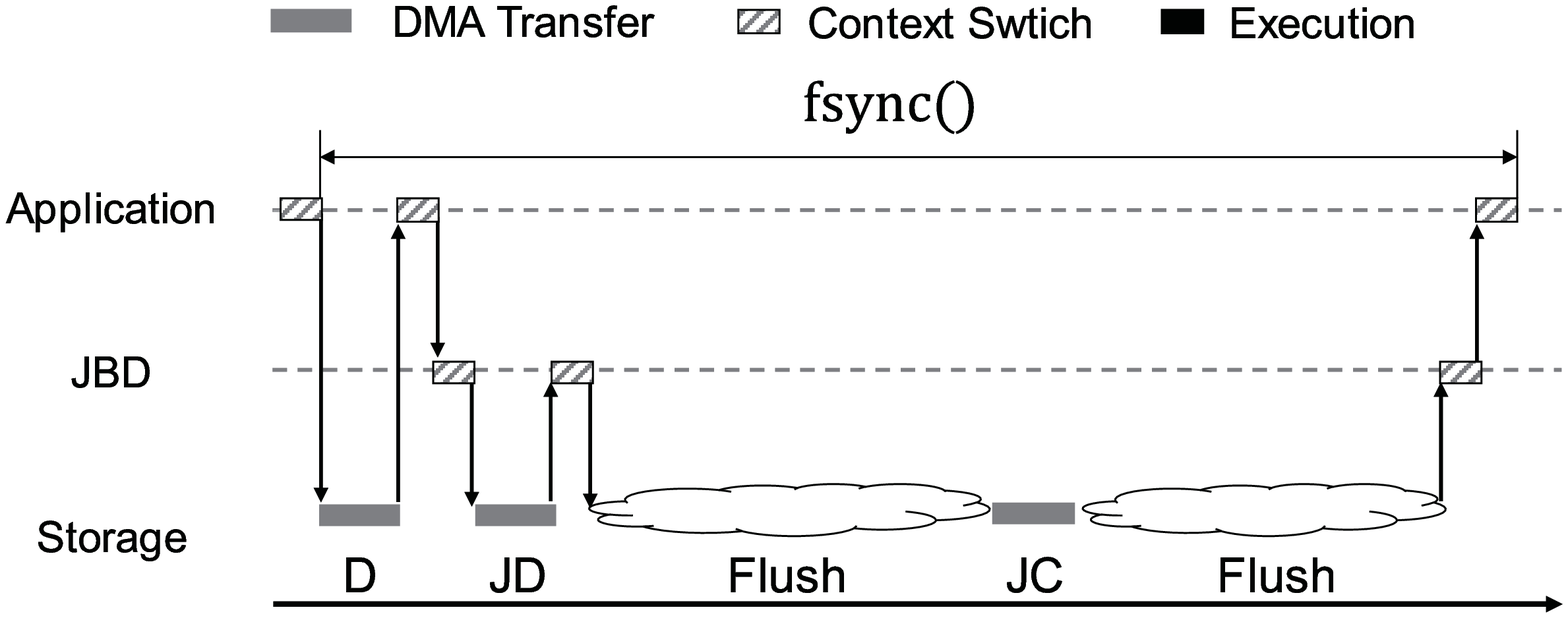}}
\\
\subfigure[\texttt{fsync()} and \texttt{fbarrier()} in BarrierFS\label{fig:BFS_cs}]{\includegraphics[width=0.45\textwidth]{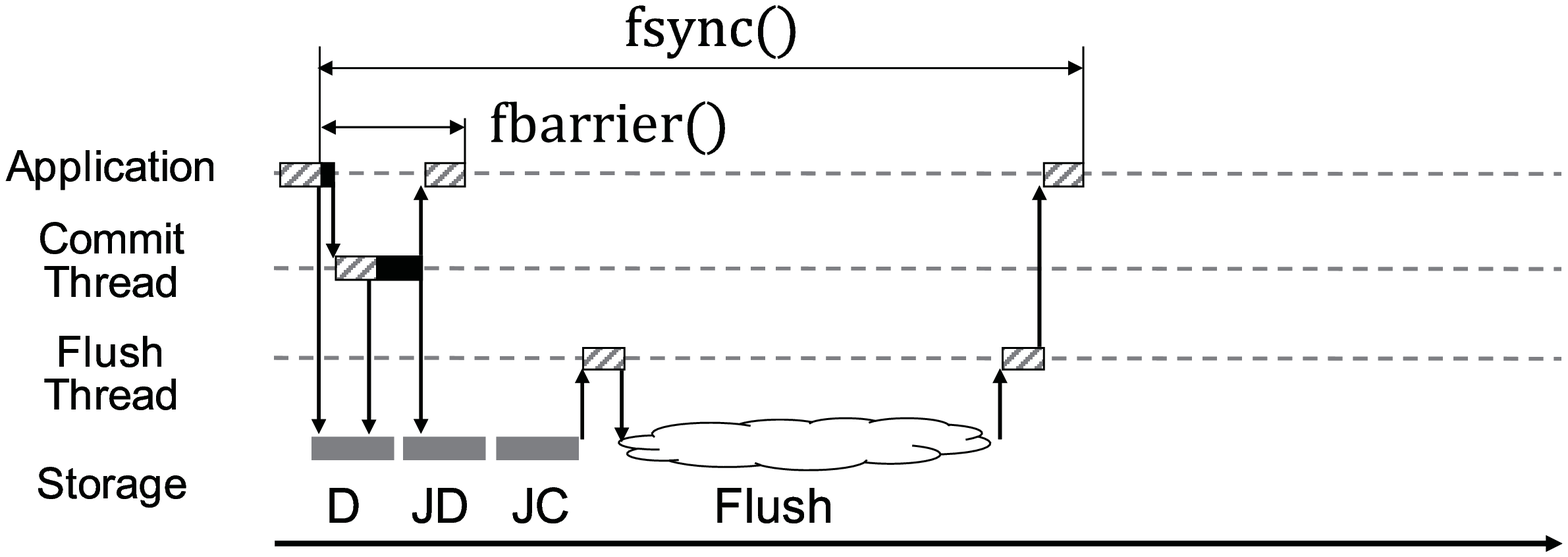}}  
\caption{\texttt{fsync()} and \texttt{fbarrier()}, D: DMA for
dirty pages, JD: DMA for journal descriptor, JC: DMA for journal commit block\label{fig:cs}}
\label{fig:tr} 
\end{figure}

\begin{comment}
%
\begin{figure}[h]
\begin{center}
\includegraphics[width=0.45\textwidth]{./figure/flow_barrierFS_fsync+fbarrier.eps}
\end{center}
\caption{\texttt{fsync()} and \texttt{fbarrier()} in BarrierFS{\color{red}:  fbarrier()가 끝나는 지점은 flush 쓰레드가 시작하는 직후가 되어야 합니다. 수정합시다.}}
\label{fig:ext4_ordered}
\end{figure}
\end{comment}
Committing a journal transaction essentially consists
of two separate tasks: dispatching write commands for $JD$ and $JC$ to 
the storage (host side) and making them durable (storage 
side). In the order preserving block device design, the host 
(the block device layer) is responsible for controlling the
dispatch order and transfer order while the storage controller takes care of 
handling the persist order. The design of order preserving
block device layer naturally supports separation of the control plane (dispatching
the write requests) and the data plane (persisting the
associated data blocks and journal transaction)
in filesystem journaling. For effective separation,
these two planes should work independently with minimum dependency.
For filesystem journaling, we allocate separate threads for dispatching 
the write requests and for making them durable: \emph{commit} thread and 
\emph{flush} thread, respectively. This mechanism is called \emph{Dual Mode Journaling}.

The commit thread is responsible for dispatching the write requests for 
$JD$ and $JC$. In BarrierFS, the commit thread tags both requests with \texttt{REQ\_ORDERED} and \texttt{REQ\_BARRIER} so that $JD$ and $JC$ are transferred and are guaranteed
to be persisted in order.
After the dispatching write request for $JC$, the commit thread
inserts the journal transaction to the committing transaction list.
In ordering guarantee (\texttt{fbarrier()}), the commit thread wakes up the 
caller. In the legacy IO stack, JBD thread interleaves the write request for $JC$ and
$JD$ with transfer-and-flush. In BarrierFS, the commit thread dispatches them
in order-preserving dispatch discipline without Wait-For-Transfer overhead
and with Wait-For-Flush overhead.

The flush thread is responsible for (i) issuing the flush command, (ii) handling
error and retry and (iii) removing the transaction from the committing transaction list. The flush thread is triggered when the $JC$ is transferred. 
If the journaling is triggered by \texttt{fbarrier()},
the flush thread removes the transaction 
from the committing transaction list and returns. It does not call flush. There
is no caller to wake up. If
the journaling is initiated by \texttt{fsync()},
the flush thread flushes the cache, removes the associated 
transaction from the committing transaction list and wakes up the 
caller. 
Via separating the control plane (commit thread) and data plane (flush thread), the commit thread can commit the following transaction 
after it is done with dispatching the write requests for preceding
journal commit. In Dual Mode journaling, there can be more than
one committing transactions in flight.

In \texttt{fsync()} or \texttt{fbarrier()}, 
the BarrierFS dispatches 
the write request for $D$ as an order-preserving request. Then,
the commit thread dispatches the
write request for $JD$ and $JC$ both  with order-preserving and barrier write.
As a result, $D$ and $JD$ form
a single epoch while $JC$ by itself forms another.  
A journal commit consists of the two epoches: $\{D, JD\}$ and $\{JC\}$.
An \texttt{fsync()} in barrierFS can be represented as in 
Eq.~\ref{eq:fsyncbfs}. Eq.~\ref{eq:fsyncbfs} also denotes the \texttt{fbarrier()}.
\begin{equation}
\underbrace{
\mbox{$D$$\rightarrow$$JD_{\mbox{\texttt{BAR}}}$ $\rightarrow$ 
$JC_{\mbox{$\texttt{BAR}$}}$}}_{\mbox{\texttt{fbarrier()}}}
\mbox{$\rightarrow$xfer$\rightarrow$flush
}
\label{eq:fsyncbfs}
\end{equation}

The benefit of Dual Mode Journaling is substantial.
In EXT4 (Fig.~\ref{fig:EXT4_cs}), an \texttt{fsync()} consists of a tandem of 
three DMA's and two flushes interleaved with context switches. 
In BarrierFS, an \texttt{fsync()} consists of single flush, three DMA's(Fig.~\ref{fig:BFS_cs}) and fewer number of context switches.
The transfer-and-flush between
$JD$ and $JC$ are completely eliminated. \texttt{fbarrier()} returns almost instantly after the commit thread dispatches the write request for $JC$. 

BarrierFS forces  journal commit if  
\texttt{fdatasync()} or \texttt{fdatabarrier()} do not find any dirty pages. 
Through this scheme, \texttt{fdatasync()} (or \texttt{fdatabarrier()}) can delimit 
an  epoch despite the absence of the dirty pages. 

\subsection{ Multi-Transaction Page Conflict}

A buffer page can belong to only one 
journal transaction at a time \cite{tweedie1998journaling}.
Blindly inserting a buffer page to the running transaction may
yield removing it from the committing transaction before it becomes durable.
We call this situation as \emph{page conflict}.
In both EXT4 and BarrierFS, when the application
thread inserts a buffer page to the running transaction, it checks if
the buffer page is being held by the committing transaction. If so, the application blocks without inserting it to the running transaction. 
When the JBD thread of EXT4 (or flush thread in BarrierFS) has made the committing transaction durable, it 
identifies the conflict pages in the committed transaction and inserts them
to the running transaction. In EXT4, there is only one committing transaction at a time. The running transaction is guaranteed to be conflict 
free
when the JBD thread resolves the page conflicts from the committed transaction.
In BarrierFS, the running transaction can conflict with more than one committing transactions, \emph{multi-transaction page conflict}. When the flush thread resolves the page conflicts from a committed transaction, 
the running transaction may
still conflict with the other committing transactions. If the running
transaction is committed prematurely with conflicted pages missing, the storage order can be compromised. 
Whenever
the flush thread resolves the page conflicts and notifies the commit thread
about its completion of persisting a transaction, 
the commit
thread has to scan all the pages in the other 
committing transactions for page conflict. To reduce the overhead
of scanning the pages, we introduce \emph{conflict-page list}. The application 
thread inserts the buffer page to the
conflict-page list  if the
buffer page is being held by one of the committing transactions.
When the flush thread 
has made the committing transaction durable, the flush thread inserts
the conflict pages to the buffer page list of the running transaction 
and removes them from the conflict-page list. The commit thread can start committing a running transaction only when  conflict-page list is
empty.

\vspace{-.3cm}
\subsection{Analysis}
\begin{figure}[hbtp]
\centering
\includegraphics[width=0.9\linewidth]{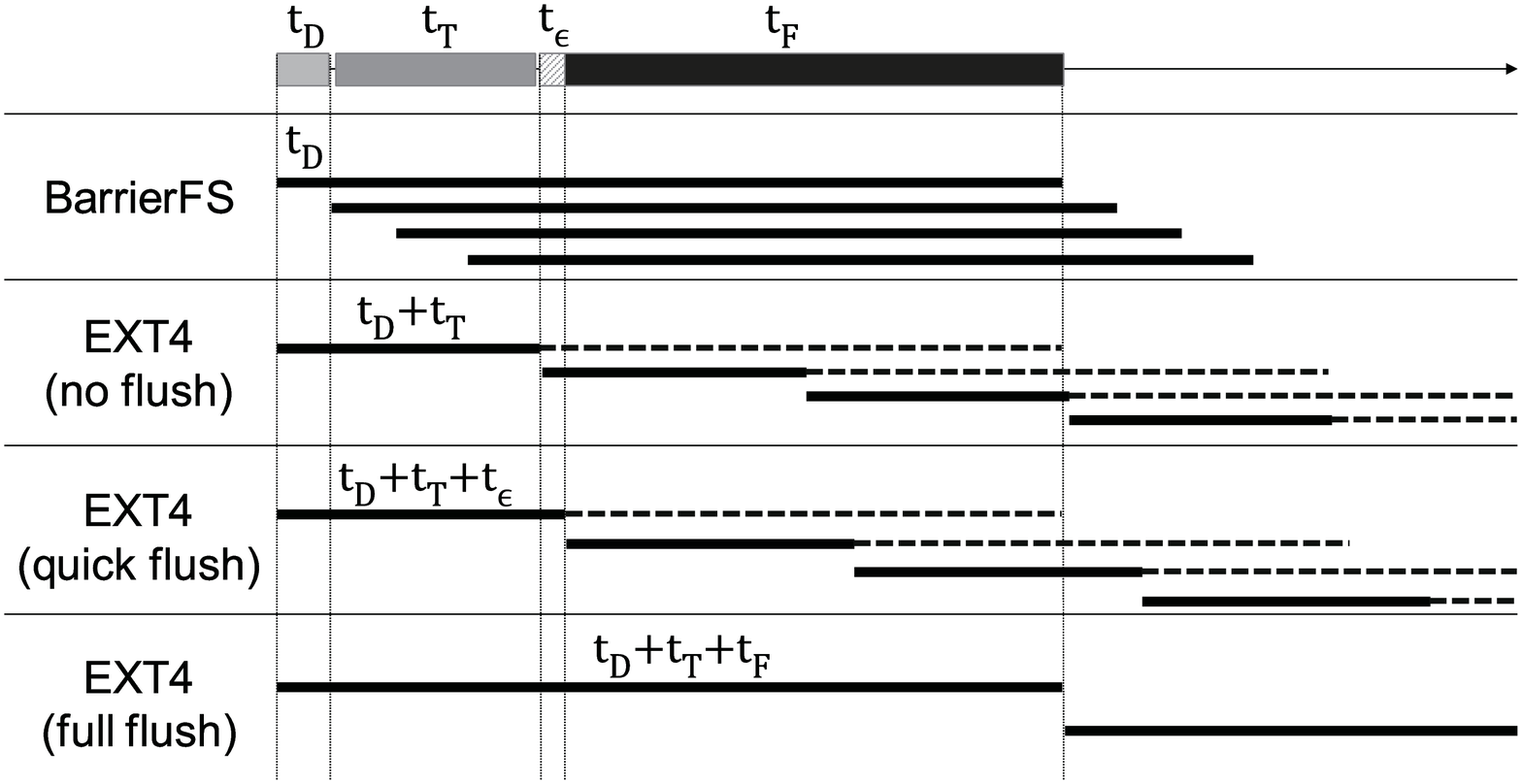}
%\caption{\texttt{fsync()} under 
%different storage order guarantee: BarrierFS, EXT4 (quick 
%flush), EXT4 (full flush), \texttt{$\Delta$D}: dispatch latency, \texttt{$\Delta$C}: transfer latency, \texttt{$\Delta\epsilon$}: flush latency in supercap SSD, \texttt{$\Delta$F}: flush latency}
\caption{\texttt{fsync()} under 
different storage order guarantee: BarrierFS, EXT4 (no flush), EXT4 (quick 
flush), EXT4 (full flush), $t_D$: dispatch latency, $t_C$: transfer latency, $t_\epsilon$: flush latency in supercap SSD, $t_F$: flush latency}
\label{fig:overhead}
\vspace{-.3cm}
\end{figure}

%
\begin{comment}
The \texttt{fbarrier()} of BarrierFS
can be represented as follows. The \texttt{fbarrier()} 
does not guarantee the durability of a transaction and thus does not require \texttt{FLUSH$|$FUA} option in writing $JC$. 
\begin{equation}
\mbox{
$D_{\mbox{\texttt{BARRIER}}}$ $\rightarrow$ $JD_{\mbox{\texttt{BARRIER}}}$ $\rightarrow$ $JC_{\mbox{\texttt{BARRIER}}}$
}
\end{equation}
\end{comment}
%
We examine how the journaling throught may vary subject to 
different
methods of journal commit: BarrierFS, EXT4 with \texttt{no-barrier} option, 
EXT4 with supercap SSD and and plain EXT4.
Fig.~\ref{fig:overhead} schematically illustrates the behaviors. With \texttt{no-barrier} mount option, filesystem does not issue flush command in \texttt{fsync()} or \texttt{fdatasync()}.
$t_D$, $t_C$ and $t_F$ denote the 
dispatch latency, transfer latency, and flush latency associated with 
committing a journal transaction, respectively. In particular, $t_\epsilon$ 
denotes the total flush latency in supercap SSD. 

With supercap SSD, EXT4 (quick flush), the journal 
commits are interleaved by $t_D + t_C + t_\epsilon$.
The host  observes the round-trip delay of the flush command and 
the associated context switch overhead, $t_\epsilon$. $t_\epsilon$ is not
negligible in Flash storage.
EXT4 with no-barrier option, EXT4 (no flush), can commit
a new transaction once all the  associated blocks are 
transferred to the storage. The journaling is interleaved by command dispatch and
DMA transfer, $t_D+t_C$.  
In BarrierFS, the commit thread keeps dispatching the journal commit
operations without waiting for the completion of the transfer.
The interval between the successive journal commit can be as small as $t_D$. 

\vspace{-0.3cm}
\section{Applications on Barrier Enabled IO stack}
\label{section:app}
\begin{comment}
Table \ref{tab:ext4vsbarrierfs} summarizes the 
number of flush operations in each filesystem. In EXT4 and BarrierFS,
there are two flushes and only one flush in \texttt{fsync()},
respectively. EXT4 filesystem issues flush after sending the dirty page
cache entries (D) and the journal descriptor (JD). The second flush is
issued after it dispatches the write request for journal commit
(JC). BarrierFS entails only one flush in \texttt{fsync()} at the very
end of the system call. 

BarrierFS exports \texttt{fbarrier()} and \texttt{fdatabarrier()} API
which does not accompany any flush operation. The application exploits
this primitive when it  requires only an ordering guarantee.

\begin{table}[t]
\caption{Ordering Guarantee in EXT4 and BarrierFS\label{tab:ext4vsbarrierfs}}
\centering
\begin{tabular}{|l|l|r|r|r|r|}
\hline
Primitives & FS & D & JD & JC\\ \hline\hline
fsync() & ext4 & - & flush & flush\\ \hline
fsync() & barrierfs & - & barrier & flush\\ \hline
fdatasync() & ext4 & flush & - & -\\ \hline
fdatasync() & barrierfs & flush & - & - \\ \hline
fbarrier() & barrierfs & - & barrier & barrier\\ \hline
fdatabarrier() & barrierfs & barrier & - & - \\ \hline
\end{tabular}
\end{table}
\end{comment}
\texttt{fsync()} accounts for dominant fraction of IO in modern applications, 
e.g.~mail server~\cite{sehgal2010evaluating} or OLTP. 90\% of IO's in the TPC-C
workload is created by \texttt{fsync()} for synchronizing
the logs to the storage ~\cite{ou2016high}. The order preserving IO stack
can significantly improve the performance in these workloads.
SQLite can be  the application which the Barrier Enabled IO stack 
benefits the most. SQLite  uses
\texttt{fdatasync()} not only to guarantee the durability of a transaction but also
to control the storage order in various occasions, e.g.~between writing the undo-log and storing the journal header and between writing updated database node and writing the commit block \cite{lee2015waldio}.
In a single insert transaction, SQLite calls \texttt{fdatasync()} four times,
three of which are
to control the storage order. We can replace them
with \texttt{fdatabarrier()}'s without compromising the durability of a
transaction.   
Some applications prefer to trade the durability and freshness of the
result with the performance and scalability of the operation
\cite{cipar2012lazybase,cui2014exploiting}. The benefit of BarrierFS can
be more than significant in these applications. One can replace all
\texttt{fsync()} and \texttt{fdatasync()} with ordering guarantee
counterparts, \texttt{fbarrier()} and \texttt{fdatabarrier()}, respectively.

\begin{comment}
\section{Implementation}
\subsection{Order Preserving Block Device Layer}
- ordered write
Ordered Write는
- barrier write
Barrier Write는

\subsection{Journaling Mode}
journaling mode: durable commit, ordering commit

\subsection{Committing transaction list}
committing transaction list

베리어 플래그가 구현된 자료구조가 어디어디인지? scsi command하고 io request는 어디에 구현되어 있나? 베리어 플래그는 Request Flag를 정의하는 enum field인  rq\_flag\_bits에 추가되어 있습니다. rq\_flag\_bits는 bio 구조체, request 구조체에서 I/O의 특성을 나타내기 위해 사용됩니다. bio 구조체는 unsigned int bi\_rw, request 구조체는 unsigned int cmd\_flags라는 변수가 각각 정의되어 있어 기존 I/O Stack에서는 해당 변수 내에 rq\_flag\_bits를 저장하지만 Order Preserving I/O Stack의 경우에는 ORDERED flag, BARRIER Flag가 추가되어 있기 때문에 bio 구조체에서는 기존에 사용하던 bi\_rw 변수를 unsigned long long으로 형변환하였고 request 구조체에서는 unsigned long long형의 cmd\_bflags를 새롭게 정의하였습니다.
\end{comment}
\vspace{-.3cm}
\section{Experiment}
\label{section:exp}
\subsection{Setup}
%우리는 본 논문에서 우리 barrier 기술의 성능 실험을 수행하였다.
%We measure the improvement of IO Barrier
We implement Barrier Enabled IO stack on three different 
platforms: smartphone (Galaxy S6, Android 5.0.2, Linux 3.10), PC server (4 cores, Linux 3.10.61) and enterprise server 
(16 cores, Linux 3.10.61). We test three storage devices:  
mobile storage (UFS 2.0, QD\footnote{QD: queue depth}=16, single
channel), 850 PRO for server (SATA 3.0, QD=32, 8 channels), 843TN for server (SATA 3.0, QD=32, 8 channels, supercap). 
We call each of these as UFS, plain-SSD and supercap-SSD, respectively. 
We implement barrier write command in UFS device. 
In plain-SSD, we introduce 5\% performance penalty to simulate the 
barrier overhead. For supercap-SSD,
we assume that there is no barrier overhead.
\subsection{Order Preserving Block Layer}
\begin{figure}[t]
\begin{center}
\includegraphics[width=0.45\textwidth]{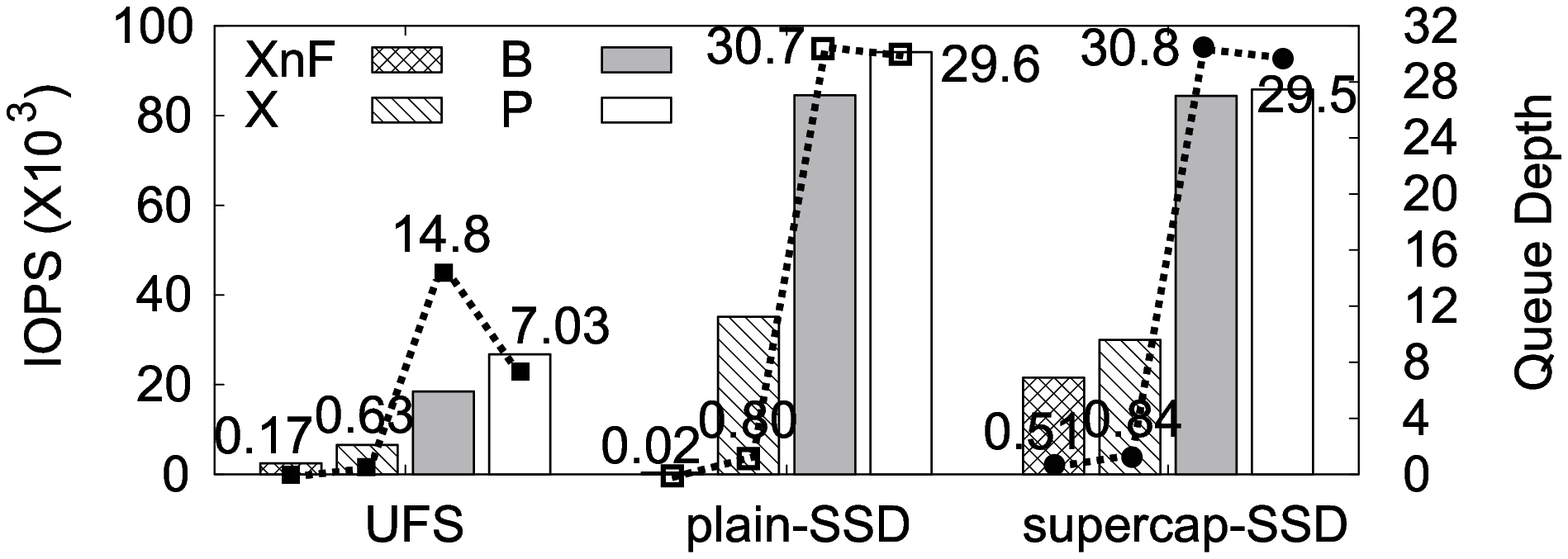}
\end{center}
\abovecaptionskip -2mm
\caption{4KB Randwom Write; XnF:  write() followed by \texttt{fdatasync()}, X:
write followed by \texttt{fdatasync()}(\texttt{no-barrier} option), B: write() followed by \texttt{fdatabarrier()}, P: Plain Buffered write()} 
\label{fig:fileIO_perf}
\vspace{-.3cm}
\end{figure}
We examine the performance of 4 KByte random write with
different ways of enforcing the storage order. 
Fig.~\ref{fig:fileIO_perf} illustrates the result. 
In  scenario `X' where `X' denotes Wait-On-Transfer, the host sends the following request after 
the data block associated with the preceding request is completely
transferred. Despite the absence of the flush overhead,
the storage devices exhibit less than 50\% of its plain buffered write
performance, the scenario `P'.
All three devices are severely underutilized. Average queue depths in all three 
devices are less than one. Wait-On-Transfer overhead in modern IO stack prohibits the host from properly exploiting the underlying Flash storage.
In  scenario `B' where `B' denotes Barrier, the IO performance increases at least by 2$\times$ 
against scenario `X'. The average queue depths reach near the maximum
in all three Flash storages. An
\texttt{fdatabarrier()} is not entirely free. We observe 1 \% to 25\% performance deficiency when it is compared against the plain buffered write. 
Plain buffered write exhibits shorter queue depth than barrier write does (Fig.~\ref{fig:fileIO_perf}).
This is because in plained buffered write, the  IO scheduler
 merges the multiple requests  and  the number of
 commands dispatched to the storage device decreases. 

\begin{figure}[t]
\centering
\subfigure[\emph{Wait-For-Transfer}, plain SSD\label{fig:WoC_SSD}]
{\includegraphics[width=0.24\textwidth]{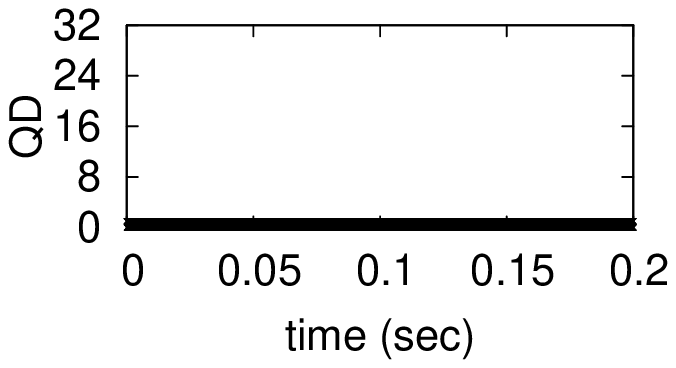}}
\hspace{-5mm} %\quad
\subfigure[\emph{Barrier}, plain SSD\label{fig:WoD_SSD}]
{\includegraphics[width=0.24\textwidth]{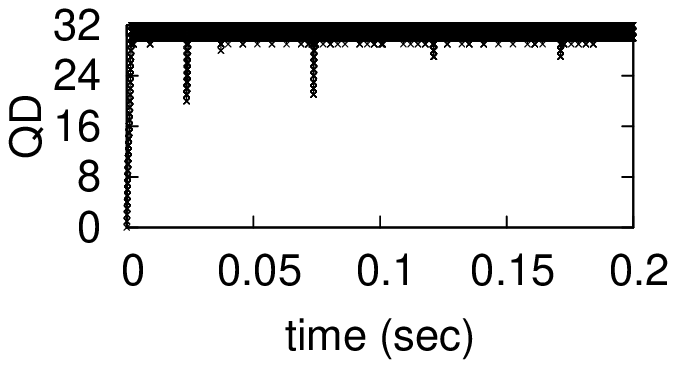}}
\\
\subfigure[\emph{Wait-For-Transfer}, UFS\label{fig:WoC}]
{\includegraphics[width=0.24\textwidth]{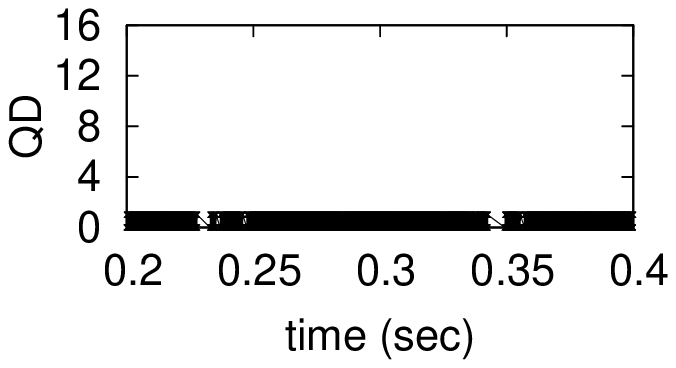}}
\hspace{-5mm} %\quad
\subfigure[\emph{Barrier}, UFS\label{fig:WoD}]
{\includegraphics[width=0.24\textwidth]{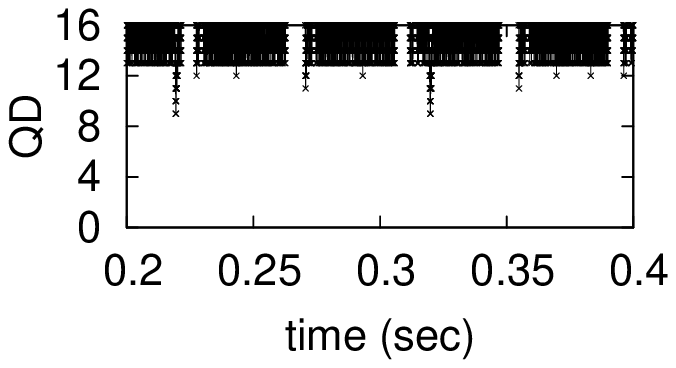}}
\caption{Queue Depth, 4KB Random Write\label{fig:QD}, \emph{Wait-For-Transfer}: write() 
followed by \texttt{fdatasync()} with \texttt{no barrier}, \emph{Barrier}: write() followed by \texttt{fdatabarrier()}}
\label{fig:QD} 
\vspace{-0.3cm}
\end{figure}

Fig.~\ref{fig:QD} is another manifestation of \texttt{fdatabarrier()}.
The storage performance is closely related to the command queue
utilization \cite{kim2015empirical}. 
When the requests are interleaved with DMA transfer, 
the queue depth never goes beyond one
(Fig.~\ref{fig:WoC_SSD} and Fig.~\ref{fig:WoC}). 
When the write request is followed by fdatabarrier(),  the queue depth grows near
to its maximum in all three storage. 
(Fig.~\ref{fig:WoD_SSD} and
Fig.~\ref{fig:WoD}).
Order preserving block layer enables the host to fully  exploit the concurrency
and the parallelism of the underlying Flash storage. 

\begin{comment}
We expect that in supercap-SSD, the flush command and the barrier
command exhibits similar latency since the SSD
controller returns both of them instantly. In supercap-SSD, the performance
of transfer-and-flush workload can be a good indicator to estimate the performance where the flush command is mechanically replaced with 
barrier command\cite{cachebarrier}.
In \emph{transfer-and-flush} workload and \emph{barrier} workload, supercap SSD exhibits 21689 IOPS and 88994 IOPS, respectively.
The Order Preserving IO stack brings 4$\times$ performance
gain against the case when the flush command is
mechanically replaced with barrier command.
We capture the
number of commands in the command queue in Fig.~\ref{fig:fileIO_QD}. 
\begin{figure}[t]
\begin{center}
%\subfigure[Orderless IO, UFS2.0\label{fig:direct_RW_QD}]{
%\includegraphics[width=0.22\textwidth, height=0.8in]{figure/direct_RW.eps}
%}
%\subfigure[Order Preserving IO, UFS2.0\label{fig:fdatabarrier_RW_QD}]{
%\includegraphics[width=0.22\textwidth, height=0.8in]%{figure/fdatabarrier_RW.eps}
%}
%\\
\subfigure[Orderless IO stack\label{fig:direct_RW_QD_SSD}]{
\includegraphics[width=0.22\textwidth,height=0.8in]{figure/direct_RW_SSD.eps}
}
\subfigure[Order Preserving IO stack\label{fig:fdatabarrier_RW_QD_SSD}]{
\includegraphics[width=0.22\textwidth,height=0.8in]{figure/fdatabarrier_RW_SSD.eps}
}
\end{center}
\abovecaptionskip -2mm
\caption{Queue Lengths in 4KB Random Write in enforcing the storage order, SSD, queue depth = 32 \label{fig:fileIO_QD}} 
%\vspace{-10pt}
\end{figure}
}
\end{comment}

\subsection{Filesystem Journaling}
\label{section:experiment_filesystem_journaling}

{\bf Latency}: 
\begin{comment}
\begin{figure*}[bhtp]
\begin{center}
\subfigure[UFS
\label{fig:gs6fsync}]{
\includegraphics[width=0.3\textwidth]{figure/CDF_Latency_GS6.eps}
}
\subfigure[plain-SSD\label{fig:CDF_850PRO}]{
\includegraphics[width=0.3\textwidth]{figure/CDF_Latency_850PRO.eps}
}
\subfigure[suercap-SSD\label{fig:CDF_843TN}]{
\includegraphics[width=0.3\textwidth]{figure/CDF_Latency_843TN.eps}
}
%\subfigure[SSD 843TN\label{fig:CDF_843TN}]{
%\includegraphics[width=0.3\textwidth]{figure/CDF_Latency_843TN.eps}
%}
\end{center}
\caption{CDF of latency: \texttt{fsync(), \texttt{fbarrier()}} \label{fig:fsynclatency}}
\end{figure*}
\end{comment}
In plain-SSD and supercap-SSD, the average \texttt{fsync()}
latency decreases by 40\%
when we use BarrierFS against when we use EXT4 (Table \ref{tab:fsync_latency}). 
UFS experiences more significant reduction in \texttt{fsync()} latency
than the SSD's do. 
The smartphone uses transactional checksum in filesystem journaling. With BarrierFS,
we can eliminate not only the transfer overhead but also the checksum overhead.
The \texttt{fsync()} latency decreases by 60\% in BarrierFS.
In supercap-SSD and UFS,
the \texttt{fsync()} latencies at 99.99$^{th}$ percentile are 30$\times$ of 
the average
\texttt{fsync()} latency(Table \ref{tab:fsync_latency}). Using BarrierFS, the tail latencies at
99.99$^{th}$ percentile decrease by 50\%, 20\% and 70\% in  
UFS, plain-SSD and supercap-SSD, respectively, against EXT4. 

\begin{table}[!h]
\begin{center}
\begin{footnotesize}
\begin{tabular}[width=0.35\textwidth]{|c||c|c||c|c||c|c|}                 \hline
& \multicolumn{2}{c||}{UFS}    & \multicolumn{2}{c||}{plain-SSD} & \multicolumn{2}{c|}{supercap-SSD}   \\ \cline{2-7}
(\%)	&	EXT4	&	BFS	&	EXT4	&	BFS	&	EXT4	&	BFS	\\	\hline
$\mu$	&	1.29 	&	0.51 	&	5.95 	&	3.52 	&	0.15 	&	0.09 	\\	\hline
Median	&	1.20 	&	0.44 	&	5.43 	&	3.01 	&	0.15 	&	0.09 	\\	\hline
99$^{th}$	&	4.15 	&	3.51 	&	11.41 	&	8.96 	&	0.16 	&	0.10 	\\	\hline
99.9$^{th}$	&	22.83 	&	9.02 	&	16.09 	&	9.30 	&	0.28 	&	0.24 	\\	\hline
99.99$^{th}$	&	33.10 	&	17.60 	&	17.26 	&	14.19 	&	4.14 	&	1.35 	\\	\hline
\end{tabular}
\end{footnotesize}
\end{center}
\abovecaptionskip -1mm
\caption{\texttt{fsync()} latency statistics (msec)}
\label{tab:fsync_latency}
\end{table}

\begin{comment}
Interestingly, the benefit of using Order Preserving IO stack
becomes more substantial in supercap SSD.
In supercap-SSD (Fig.~\ref{fig:CDF_843TN}), 
\texttt{fsync()} latency decreases by {\color{blue}8/10} when we use BarrierFS
instead of using EXT4 with no-barrier option ({\color{blue} 0.10ms vs.~0.13ms}).
In \texttt{fsync()}, BarrierFS removes context switch and DMA transfer overhead whereas 
EXT4 with no-barrier option eliminates the flush overhead.
Context switch overhead and DMA transfer overhead 
account for relatively larger fraction of \texttt{fsync()} latency 
in supercap SSD.  
The benefit of eliminating the latency of DMA transfers and context switches outweighs the benefit of eliminating the flush operation.
\end{comment}
\begin{comment}
\begin{figure}[t]
\begin{center}
 \includegraphics[width=0.4\textwidth]{figure/RW_CS.eps}
\end{center}
\abovecaptionskip -2mm
\caption{Average Number of Context Switches in Writing Single 4 KByte Block} 
\label{fig:RW_CS}
\end{figure}
\end{comment}

\begin{figure}[t]
\begin{center}
\includegraphics[width=0.45\textwidth]{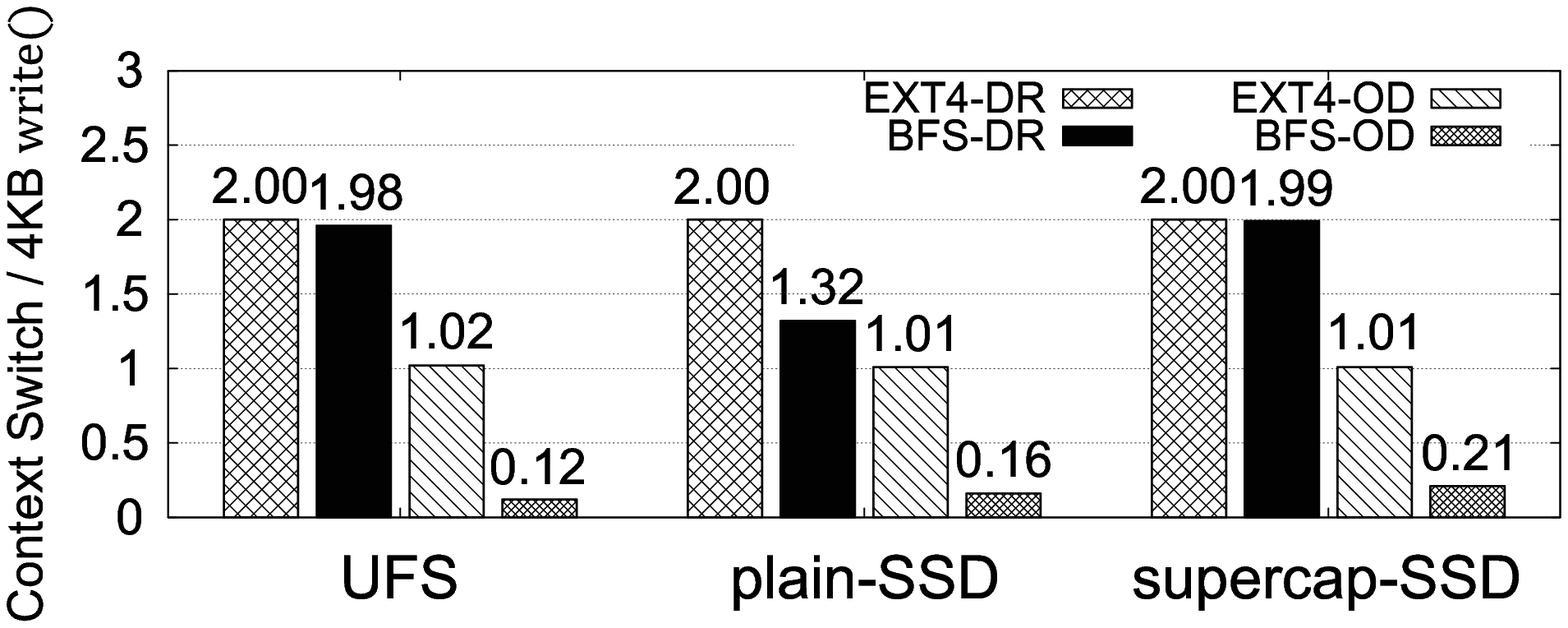}
\end{center}
\abovecaptionskip -2mm
\caption{Average Number of Context Switches per \texttt{fsync()}/\texttt{fbarrier()}, 4 KByte \texttt{write()} followed by \texttt{fsync()} or \texttt{fbarrier()}, EXT4-DR:  \texttt{fsync()}, 
BFS-DR: \texttt{fsync()}, EXT-OD: \texttt{fsync()} with \texttt{no-barrier},  
BFS-OD: \texttt{fbarrier()}} 
\label{fig:RW_CS}
\vspace{-0.3cm}
\end{figure}

{\bf Context Switches:} 
We examine the number of application level context
switches in various modes of journaling.
Fig.~\ref{fig:RW_CS} illustrates the result. 
In EXT4-DR, \texttt{fsync()} wakes up the caller twice; 
after DMA transfer of $D$ completes and after the journal transaction
is made durable. This applies to all three Flash storages.
In BarrierFS, \texttt{fsync()} wakes up the caller only once; after the
transaction is made durable. 
In UFS and supercap SSD, \texttt{fsync()} of BFS-DR wakes up the caller 
twice in entirely
different reasons. In UFS and supercap-SSD, the interval between the successive write requests are much smaller than
the timer interrupt interval due to small flush latency. As a result, \texttt{write()} requests rarely update
the time fields of the inode and \texttt{fsync()} 
becomes an \texttt{fdatasync()}. 
\texttt{fdatasync()} wakes up
the caller twice in BarrierFS; after  transferring $D$  and after flush completes.
The plain-SSD uses TLC flash. The  interval between the successive write()'s
can be longer than the timer interrupt interval.  In plain-SSD, \texttt{fsync()} occasionally commits journal transaction and the average number of context switches becomes less than two in BFS-DR for plain-SSD. 

BFS-OD manifests the benefits of BarrierFS. 
The \texttt{fbarrier()} rarely finds updated metadata since
it returns quickly. Most \texttt{fbarrier()} calls are serviced as
\texttt{fdatabarrier()}. \texttt{fdatabarrier()} does not block the caller
and it does not release CPU voluntarily. The number of context switches in
\texttt{fbarrier()} is much smaller than EXT4-OD.
BarrierFS significant improves
the context switch overhead against EXT4.

\begin{comment}
%%%%%%%%%%%%%%%%%%%%%%%%%%%%%%%%%%% RW Context Switch  %%%%%%%%%%%%%%%%%%%%%%%%%%%%%%%%%%%%%
\begin{table}[t]
\caption{Random Write context switch (Context switch / 4KByte IO)\label{conSW_table}}
\centering
\begin{tabular}{|l|r|r|r|}
\hline
\bfseries Mode & \bfseries UFS & \bfseries SSD\\ \hline
EXT4-DR				&	2.00 & 2.00\\ \hline
BFS-DR		&	1.08 & 1.92\\ \hline
EXT4-OD		&	1.02 & 1.01\\ \hline
BFS-OD	&	0.12 & 0.16\\ \hline
\end{tabular}
\end{table}

%%%%%%%%%%%%%%%%%%%%%%%%%%%%%%%%%%%%%%%%%%%%%%%%%%%%%%%%%%%%%%%%%%%%%%%%%%%%%%%%%%%%%%%%%%%%%%%%%%%%%%
\end{comment}

\begin{figure}[t]
\begin{center}
\subfigure[Durability Guarantee\label{fig:dg}]{
\includegraphics[width=0.23\textwidth]{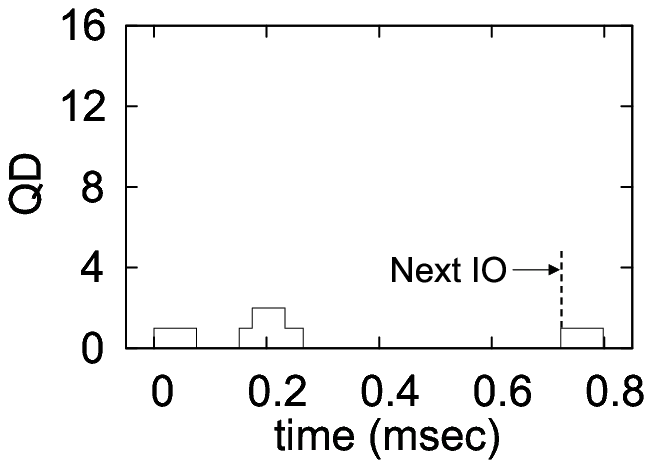}
}
\hspace{-3mm} %\quad
\subfigure[Ordering Guarantee\label{fig:og}]{
\includegraphics[width=0.23\textwidth]{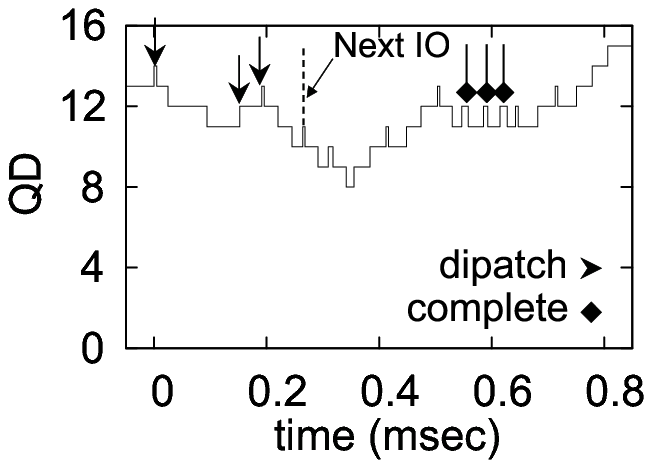}
}
\end{center}
\abovecaptionskip -2mm
\caption{Queue Depth Chanages in BarrierFS: \texttt{write()} followed by \texttt{fsync()} vs. \texttt{write()} followed by \texttt{fbarrier()}
\label{fig:qd}}
\vspace{-.5cm}
\end{figure}
%%%%%%%%%%%%%%%%%%%%%%%%%%%%%%%%%%%%%%%%%%%%%%%%%%%%%%%%%%%%%%%%%%%%%%%%%%%%%%%%%%%%%%%%%%%%%%%%%%%%%%%%%%%%%%%%%%%%%%%%%%%%%
%
%We show the queue depth in writing one 4 KByte block. In EXT4, the queue
%depth does not grow beyond one in both Durability Guarantee
%model(\texttt{EXT-DR}) and Ordering Guarantee (\texttt{EXT-OD}) model. 
{\bf Command Queue Utilization:}
In BarrierFS, \texttt{fsync()} drives the queue upto two (Fig.~\ref{fig:dg}).
Theoretically, it can drive the queue depth upto three because the host can dispatches the  write requests for $D$, $JD$ and $JC$, in tandem.
According to our instrumentation, there exists 160 $\mu
sec$ context switch interval between the application thread and the commit
thread. It takes approximately 70$\mu sec$ to transfer a 4
KByte block from the host to device cache. The command from the
application thread is serviced before the commit thread dispatches the 
command for writing $JD$.  
In \texttt{fbarrier()}, BarrierFS successfully saturates the command 
queue (Fig.~\ref{fig:og}). The queue depth increases to
fifteen. 

{\bf Throughput:}
\begin{figure}[b]
\begin{center}
\subfigure[plain-SSD\label{fig:DWSL_DR_DELL2}]{
\includegraphics[width=0.22\textwidth]{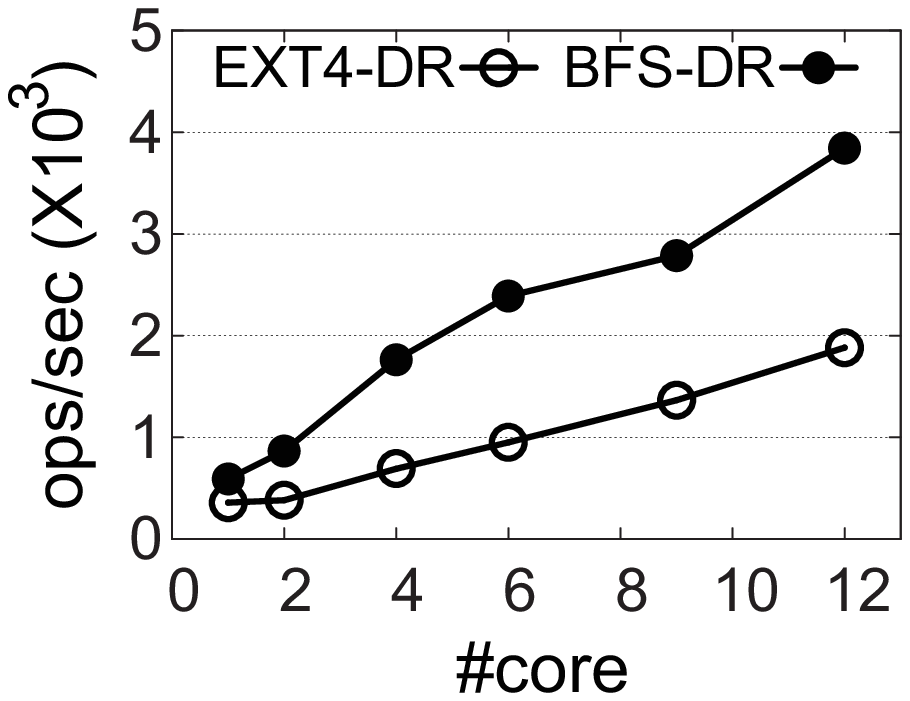}
}
\hspace{-1mm} %\quad
\subfigure[supercap-SSD\label{fig:DWSL_OD_DELL2}]{
\includegraphics[width=0.22\textwidth]{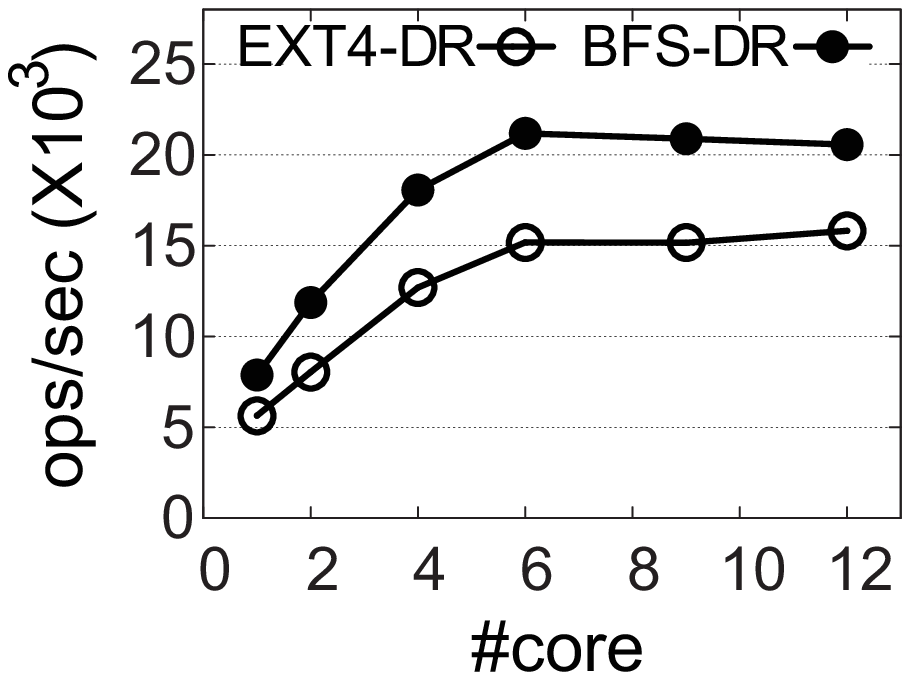}
}
\end{center}
\abovecaptionskip -2mm
\caption{fxmark: scalability of filesystem journaling\label{fig:DWSL_DELL}}
\end{figure}
We examine the throughput of filesystem journaling under varying  number of
CPU cores. We use modified DWSL workload in
\texttt{fxmark} \cite{min2016manycore}. In DWSL workload, each thread
performs 4 Kbyte allocating write followed by 
\texttt{fsync()}. Each thread operates on its own file. Each thread writes
total 1 GByte. BarrierFS exhibits much more scalable
behavior than EXT4 (Fig.~\ref{fig:DWSL_DELL}). In plain-SSD,  
BarrierFS exhibits 2$\times$ performance against
EXT4 in all numbers of cores (Fig.~\ref{fig:DWSL_DR_DELL2}).
In supercap-SSD, the performance saturates with six cores in both EXT4
and BarrierFS.
BarrierFS exhibits 1.3$\times$ journaling throughput against EXT4 at the full 
throttle (Fig.~\ref{fig:DWSL_OD_DELL2}).

\vspace{-.1cm}

\subsection{Mobile Workload: SQLite}
\begin{figure}[h]
\begin{center} 
\subfigure[UFS\label{fig:gs6_SQLite_perf}]{
\includegraphics[width=0.23\textwidth]{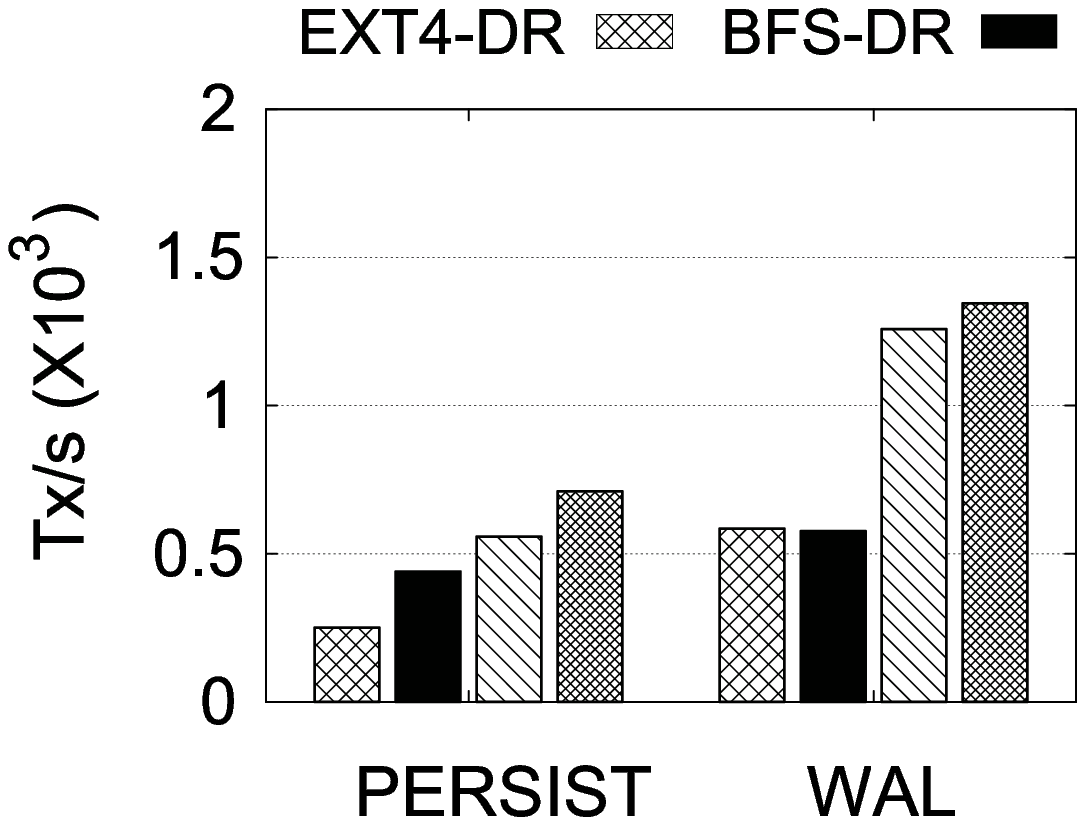}
}
\hspace{-3mm}
\subfigure[plain-SSD\label{fig:PC_SQLite_perf}]{
\includegraphics[width=0.23\textwidth]{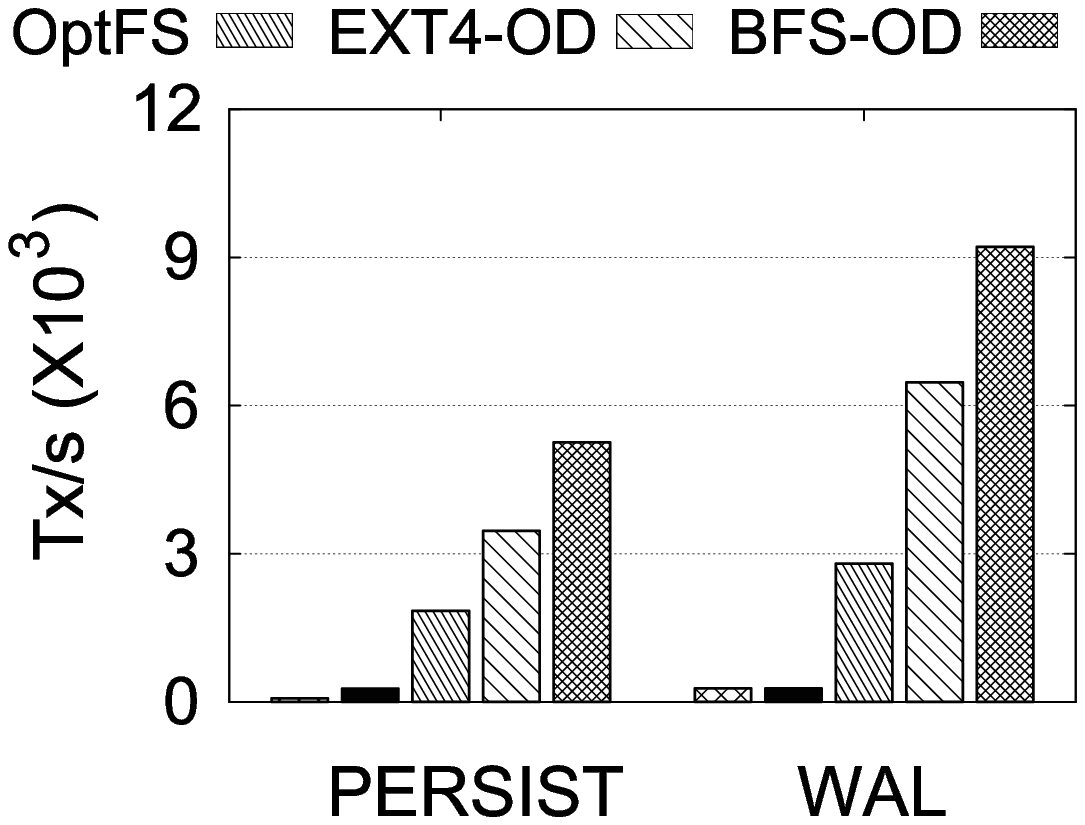}
}
\end{center}
\abovecaptionskip -2mm
\caption{SQLite Performance: inserts/sec (100,000 inserts)\label{fig:SQLite_IOPS}}
\vspace{-.5cm}
\end{figure}

In mobile storage, 
BarrierFS achieves 75\% performance improvement against EXT4 in default PERSIST journal 
mode under durability guarantee (Fig.~\ref{fig:SQLite_IOPS}).
We replace first three
\texttt{fdatasync()}'s with \texttt{fdatabarrier()}'s among all four
\texttt{fdatasync()}'s in a transaction. We keep the last
\texttt{fdatasync()} for the durability of a transaction.
In Ordering guarantee, we replace all four \texttt{fdatasync()}'s with
\texttt{fdatabarrier()}'s. 
When we remove the durability
requirement, the performance increases by 2.8$\times$ 
 in PERSIST mode against the baseline EXT4.
 In WAL mode, SQLite issues \texttt{fdatasync()} once in
every commit and there is not much room for improvement for BarrierFS.

The benefit of eliminating the Transfer-and-flush is more significant
as the storage has higher degree of parallelism and slow Flash device.
In plain-SSD, SQLite exhibits 73$\times$ performance gain in BFS-OD
against baseline EXT4-DR.
\begin{comment}
The benefit of BarrierFS is more significant in server storage. 
In Durability guarantee, BarrierFS achieves 
4$\times$ performance gain against EXT4 in PERSIST mode. In Ordering
guarantee, BarrierFS achieves 76$\times$  
performance improvement against EXT4 baseline (72 ins/sec vs.~5535
ins/sec) in PERSIST mode. In WAL mode, with ordering guarantee,
BarrierFS achieves 35$\times$ performance gain against the baseline (274
ins/sec vs.~9700 ins/sec).   
\end{comment}
%

\begin{comment}
\begin{figure}[t]
\begin{center}
%\subfigure[EXT4\label{fig:SQLite_QDa}]{
%\includegraphics[width=0.45\textwidth]{figure/fdatasync_p.eps}
% }
%\\
%\subfigure[BarrierFS-DR\label{fig:SQLite_QDb}]{
%\includegraphics[width=0.45\textwidth]{figure/fdatabarrier_p.eps}
% }
%\\ls
%\subfigure[BarrierFS\label{fig:SQLite_QDc}]{
% }
{
\includegraphics[width=0.45\textwidth]{figure/insert_QD.eps}
}
\end{center}
\abovecaptionskip -2mm
\caption{Queue depth changes: a single \texttt{insert} transaction
in SQLite (PERSIST Mode): EXT4 vs.~BarrierFS(Galaxy S6, UFS 2.0)
\label{fig:SQLite_QD}}
\end{figure}
%

Fig.~\ref{fig:SQLite_QD} illustrates the queue depth behavior of SQLite 
under EXT4 and BarrierFS, respectively. In stock SQLite, 
it takes nearly 4 msec to complete a single transaction and the queue
depth never goes up beyond three. In BarrierFS, SQLite is
modified to enforce the storage order as described in section \ref{section:app}. The command queue is well exploited. It takes 
approximately 1.0 msec to finish the transaction.
\end{comment}
\vspace{-.1cm}
\subsection{Server Workload}
We run two workloads: \texttt{varmail} workload in 
FILEBENCH~\cite{wilson2008new} and  OLTP-insert workloads from
sysbench~\cite{sysbench}. Sysbench is database workload and uses MySQL~\cite{mysql2007mysql}. \texttt{varmail} is metadata
intensive workload. 
We also test OptFS \cite{optfs2013}. We use \texttt{osync()} in OptFS. 
%
\begin{comment}
\begin{figure}[h]
\begin{center}
\includegraphics[width=0.45\textwidth]{figure/benchmark.eps}
\end{center}
\abovecaptionskip -2mm
\caption{Performance for Server Workloads, 850 PRO {\color{red}843TN 결과 Appendix 추가}} 
\label{fig:x86_filebench_perf}
\end{figure}
\end{comment}
%
\begin{comment}
\begin{figure}[h]
\begin{center}
\includegraphics[width=0.45\textwidth]{figure/Performance_Server_Workloads_850_843.eps}
\end{center}
\abovecaptionskip -2mm
\caption{Performance for Server Workloads, 850PRO and 843TN} 
\label{fig:x86_filebench_perf}
\end{figure}
\end{comment}
%
\begin{comment}
\begin{figure}[h]
\begin{center}
\subfigure[plain-SSD\label{fig:x86_filebench_850}]{
\includegraphics[width=0.21\textwidth]{figure/Performance_Server_Workloads_850_8.eps}
}
\quad
\subfigure[supercap-SSD\label{fig:x86_filebench_843}]{
\includegraphics[width=0.21\textwidth]{figure/Performance_Server_Workloads_843_8.eps}
}
\end{center}
\abovecaptionskip -10mm
\caption{Performance for Server Workloads, 850PRO and 843TN}
\label{fig:x86_filebench_perf}
\end{figure}
\end{comment}
%
\begin{figure}[b]
\begin{center}
\includegraphics[width=0.45\textwidth]{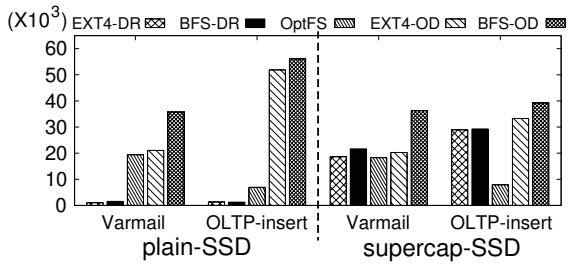}
\end{center}
\abovecaptionskip -2mm
\caption{Performance for Server Workloads, Filebench: Varmail(ops/s), Sysbench: OLTP-insert(Tx/s)
\label{fig:x86_filebench_perf}}
\vspace{-.5cm}
\end{figure}

We perform two sets of experiments. First, we leave the application
intact and replace  the EXT4 with BarrierFS (EXT4-DR and BFS-DR). 
We compare the \texttt{fsync()} performance between BarrierFS and EXT4.
The second set of experiment is for ordering guarantee. In EXT4, we use \texttt{nobarrier} mount option. In BarrierFS, we replace \texttt{fsync()} with \texttt{fbarrier()}. 
Fig.~\ref{fig:x86_filebench_perf} illustrates the result. 

In plain-SSD, BFS-DR brings 60\% performance gain against EXT4-DR in 
\texttt{varmail} workload.
This is due to the more efficient implementation of \texttt{fsync()} in BarrierFS. 
The benefit of BarrierFS manifests itself when we relax the durability guarantee.
The \texttt{varmail} 
workload is known for its heavy \texttt{fsync()} traffic. In EXT4-OD, the journal 
commit operations
are interleaved by DMA transfer latency. In BFS-OD, the journal commit
operations are interleaved by the dispatch latency.
The Dual mode journal can significantly improve the
journaling throughput via increasing the concurrency in journal commit.
With ordering guarantee, BarrierFS achieves 80\% performance gain against
EXT4 with no-barrier option.

In MySQL, BFS-OD prevails EXT4-OD, by 12\%. The performance increases 43$\times$ when we replace the 
\texttt{fsync()} of EXT4 with \texttt{fbarrier()}.

{\bf Notes on OptFS:} In SQLite (Fig.~\ref{fig:PC_SQLite_perf}), 
varmail and
MySQL (Fig.~\ref{fig:x86_filebench_perf}), we observe that OptFS 
does not show as good performance
in Flash storage 
as it does in the rotating media \cite{optfs2013}.
OptFS is elaborately designed to  reduce the seek overhead
inherent in Ordered mode journaling of EXT4. OptFS achieves this objective via two innovations: via flushing larger number of 
transactions together and via  selectively journaling the  data blocks.
Benefit of eliminating a seek overhead is marginal 
for Flash storage. Due to this reason, in varmail workload which
rarely entails selective data mode journaling, 
OptFS and EXT4-OD exhibit similar performance in Flash storage(Fig.~\ref{fig:x86_filebench_perf}). The selective data mode journaling  increases the amount of pages to scan for 
\texttt{osync()}, only a few of which 
can  be dispatched to the storage. The selective data mode journaling
can negatively interfere with the \texttt{osync()} especially when the
underlying storage has short latency. In \cite{optfs2013}, MySQL performance
decreases to one thirds in OptFS against EXT4-OD and the selective data mode
journaling has been designated as its prime cause. Our MySQL workload creates 
even larger amount of selective data journaling
and the performance of OptFS corresponds to one eights of
that of EXT-OD under MySQL workload (Fig.~\ref{fig:x86_filebench_perf}).

\vspace{-.2cm}

\section{Related Work}
\label{section:rel}
 
OptFS \cite{optfs2013} is the closest work
of our sort; they proposed a new journaling primitive \texttt{osync()} which returns without persisting the journaling transaction and yet which guarantees that the write requests associated with journal commits are stored in order. OptFS does not
provide the filesystem primitive that corresponds to \texttt{fdatabarrier()} in our Barrier Enabled IO stack. \texttt{osync()} still relies on Wait-On-Transfer in enforcing 
the storage order.
Featherstitch\cite{gfs2007} propose a programming model to
specify the set of requests that can be scheduled together, \texttt{patchgroup} and the ordering dependency between them
\texttt{pg\_depend()}. 
While xsyncfs~\cite{nightingale2006} successfully mitigates the overhead of \texttt{fsync()}, xsyncfs maintains complex causal dependencies among buffered updates.
An order preserving block device layer can make the implementation of xsyncfs much simpler.
NoFS (no order file system)~\cite{nofs2012} introduces ``backpointer'' to entirely eliminate the transfer-and-flush ordering requirement in the file system.
However, it does not support atomic transactions.

A few works proposed to use multiple running transaction or 
multiple committing transaction to circumvent the transfer-and-flush
overhead in filesystem journaling~\cite{IceFS2014,SpanFS2015,ccfs}, to improve journaling performance or to isolate errors.
IceFS~\cite{IceFS2014} allocates separate running transactions for each container.
SpanFS~\cite{SpanFS2015} splits a journal region into multiple partitions and allocates committing transactions for each partition.
CCFS~\cite{ccfs} allocates separate running transactions for individual threads.
These systems, where each journaling session still relies on the transfer-and-flush mechanism in enforcing the intra- and inter-transaction storage orders, are complementary to our work. 

A number of file systems provide a multi-block atomic write feature~\cite{Dabek2001CFS, F2FS2015, Msync2013, AdvFS2015} to relieve applications from the overhead of logging and journaling.
These file systems internally use the transfer-and-flush mechanism to enforce the storage order between write requests for data blocks and associated metadata.
An order preserving block device can effectively mitigate overheads incurred when enforcing the storage order in these file systems.

\vspace{-.2cm}
\section{Conclusion}
\label{section:conc}
In this work, we develop an Barrier Enabled IO stack to address the transfer-and-flush overhead inherent in the legacy IO stack. Barrier Enabled IO stack effectively eliminates the transfer-and-flush overhead associated with controlling the storage order and is successful in fully exploiting the underlying Flash storage. We like to
conclude this paper with two important observations. First, ``cache barrier'' is a necessity 
than a luxury. ``cache barrier'' is an essential tool for the host to control the 
persist order which has not been possible before. 
Currently, cache barrier command is only available in the 
standard command set for mobile storage. Given its implication on IO stack, it 
should 
be available in all range of the storage device ranging from the mobile
storage to the high performance Flash storage with supercap. 
Second, eliminating a ``Wait-On-Transfer'' overhead is not an option. It
blocks the caller and stalls the command queue leaving the
storage device being severely underutilized. As the storage latency becomes shorter,
the relative cost of ``Wait-On-Transfer'' can become more significant.

Despite all the preceding sophisticated techniques to optimize the legacy IO stack for Flash storage, we carefully argue that the IO stack is still 
fundamentally driven by the old legacy that the host cannot control the persist order.
This work shows how the IO stack can evolve when the persist order can be controlled and
its substantial benefit. We hope that this work serves as a possible basis for the 
future IO stack in the era of Flash storage.

{\footnotesize \bibliographystyle{acm}
\bibliography{ref}

\end{document}